\documentclass{acmsiggraph}                     

\usepackage[scaled=.92]{helvet}
\usepackage{times}

\usepackage{graphicx}

\usepackage{parskip}

\usepackage[labelfont=bf,textfont=it]{caption}

\usepackage{subfigure}
\usepackage{comment}
\usepackage{amsmath}

\onlineid{0}

\title{Ray--Based Reflectance Model for Diffraction}

\author{Tom Cuypers$^{(1,2)}$ \hspace{0.9cm} Se Baek Oh$^2$ \hspace{0.9cm} Tom Haber$^1$ \hspace{0.9cm} Philippe Bekaert$^1$ \hspace{0.9cm} Ramesh Raskar$^2$ \\ \vspace{-0.3cm} \\
	      $^1$ Expertise Centre for Digital Media / Hasselt University \hspace{0.8cm} $^{2}$ MIT
}  


\begin{document}


\maketitle


\begin{abstract}
We present a novel method of simulating wave effects in graphics using ray--based renderers with a new function: the Wave BSDF (Bidirectional Scattering Distribution Function). Reflections from neighboring surface patches represented by local BSDFs are mutually independent. However, in many surfaces with wavelength-scale microstructures, interference and diffraction requires a joint analysis of reflected wavefronts from neighboring patches. We demonstrate a simple method to compute the BSDF for the entire microstructure, which can be used independently for each patch. This allows us to use traditional ray--based rendering pipelines to synthesize wave effects of light and sound. We exploit the Wigner Distribution Function (WDF) to create transmissive, reflective, and emissive BSDFs for various diffraction phenomena in a physically accurate way. In contrast to previous methods for computing interference, we circumvent the need to explicitly keep track of the phase of the wave by using BSDFs that include positive as well as negative coefficients. We describe and compare the theory in relation to well understood concepts in rendering and demonstrate a straightforward implementation. In conjunction with standard raytracers, such as PBRT, we demonstrate wave effects for a range of scenarios such as multi--bounce diffraction materials, holograms and reflection of high frequency surfaces.


\end{abstract}




\section{Introduction}

Diffraction is a common phenomenon in nature when dealing with small scale occluders. It can be observed on animals, such as feathers and butterfly wings, and man-made objects like rainbow holograms. In acoustics, the effect of diffraction is even more significant due to the much longer wavelength of sound. In order to simulate effects such as interference and diffraction within a ray based framework, the phase of light or sound waves needs to be integrated into those methods.

\begin{figure}[t!]
\centerline{\includegraphics[width=0.7\columnwidth]{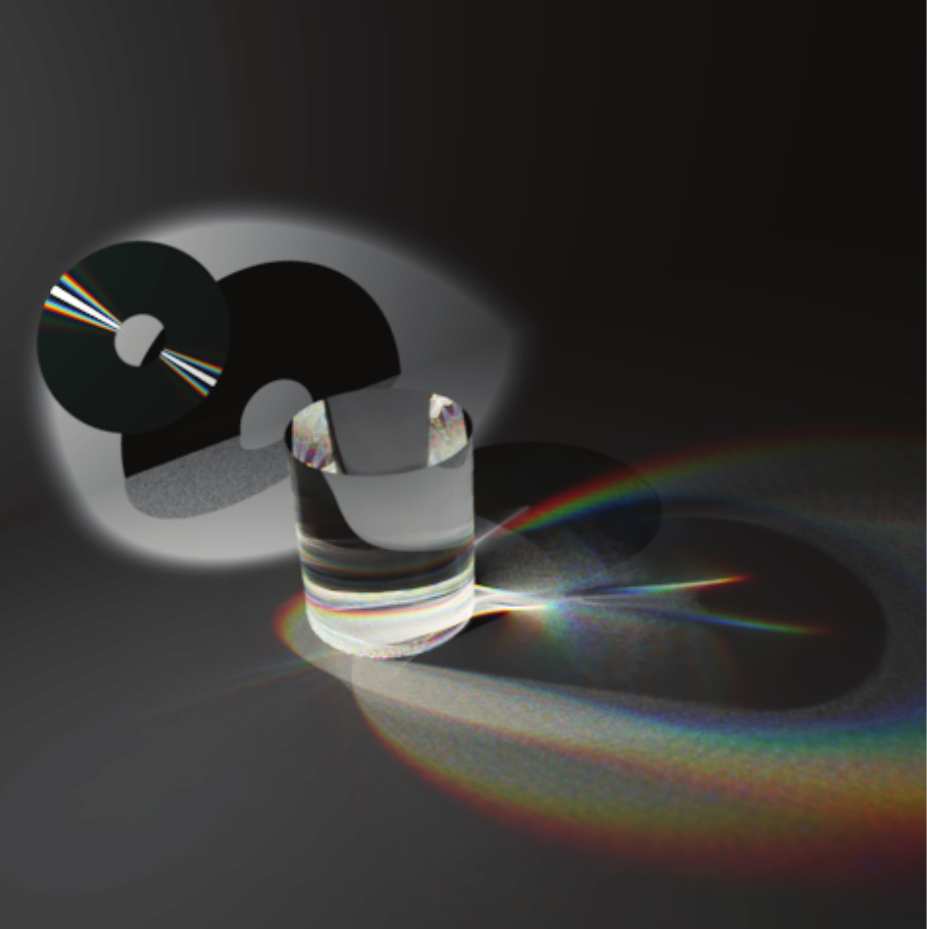}}
\caption{We generalize the rendering equation and the BSDF to simulate wave phenomena using an unmodified PBRT rendering framework. The new Wave BSDF behaves like a local scattering function, creates interference globally, and allows easy integration into traditional ray based methods.}
\label{fig:teaser}
\end{figure}

\begin{figure*}[t!]
\begin{center}
 \captionsetup{type=table}
 \centerline{\includegraphics[width=\textwidth]{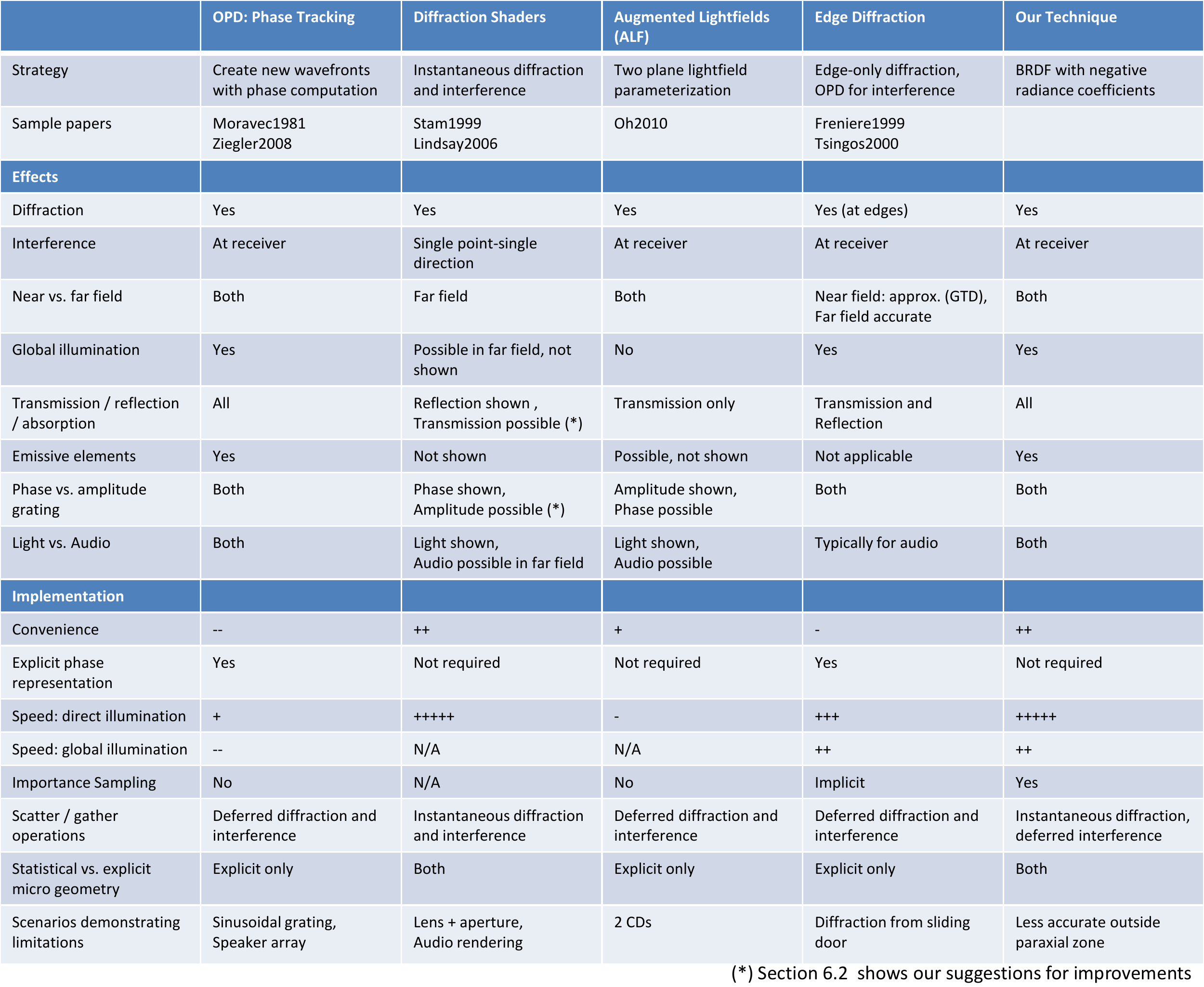}}
 \caption{Overview of different rendering strategies for diffraction and interference. Our new technique addresses all of the required features with high efficiency and simplicity. OPD is the only technique that encompasses all of the effects that WBSDF simulates, but WBSDF is more efficient and easier to use with standard rendering pipelines. }
 \label{tb:overview_rendering_methods}
\end{center}

\end{figure*}


We introduce a novel method for creating Bidirectional Scattering Distribution Functions (BSDFs), which efficiently simulate diffraction and interference in ray-based frameworks. The reflected or scattered radiance of a ray indirectly and independently encodes the mutual phase information among the rays, conveniently allowing for interference after multiple bounces. Our BSDFs, derived from the Wigner Distribution Function (WDF) in wave optics,  abstract away the complexity of phase calculations. Traditional ray--based renderers, without modifications, can directly use these WBSDFs. The implementation of the WBSDFs does not require a reader to fully understand the theory of the WBSDF derivation. A reader who is interested solely in implementing our new shaders can proceed directly to section~\ref{sec:convertBRDF}. Specific technical contributions are as follows:
\begin{enumerate}
\setlength\itemsep{-0.5\parsep}
  \item A method to compute WBSDF from microstructures
  \item Formulation of the rendering equation to simulate diffraction and interference of light
  \item A practical way to support interference and diffraction in global illumination with importance sampling
  \item Compatibility demonstration with ray--based renderers by creating a material plugin for PBRT~\cite{PBRT04}
  \item Application of the new model to simulate rainbow holograms and camera optics
  \item Comparison to other diffraction techniques and suggestions for further improvement
  \item An extension towards sound rendering (see appendix)
\end{enumerate}

\subsection{Related Work}

\textbf{Light Propagation in Optics:} In wave optics, light is described as an electromagnetic field with amplitude and phase. The Huygens--Fresnel principle is often used to represent wave propagation, which is a convolution of point scatterers~\cite{Goodman2005}. In contrast, geometrical optics treats light as a collection of rays. Among the extensive efforts to connect wave and ray optics~\cite{Wolf1978}, notable ones are the generalized radiance proposed by Walther~\shortcite{Walther1973} and the Wigner Distribution Function~\cite{Bastiaans1977}, where light is described in terms of local spatial frequency. Although the generalized radiance or the WDF can be negative, it exhibits convenient properties that explains diffraction rigorously~\cite{Bastiaans1997}. \\

\noindent\textbf{Traditional Light Propagation in Graphics:} Ray--based rendering systems, e.g., ray tracing~\cite{Whitted1980}, are popular for rendering photorealistic images in computer graphics due to their simplicity and efficiency. They are particularly convenient for simulating reflection and refraction. In addition, in combination with global illumination techniques such as photon mapping~\cite{Jensen1996,Jensen1998}, caustics and indirect light can be constructed. The idea of using negative light in the rendering equation has been proposed for visibility calculations~\cite{Dachsbacher2007}, but not for interference calculations. \\

\noindent\textbf{Wave--based Image Rendering:}  Moravec proposed using a wave model to render complex light transport efficiently~\shortcite{Moravec1981}. Stam implemented a diffraction shader based on the Kirchhoff integral~\shortcite{Stam1999} for random or periodic patterns. Other variations of diffraction based BRDFs were created for rendering specific types of materials~\cite{Sun2000,Sun2006}. Some other examples are based on the Huygens--Fresnel principle~\cite{Lindsay2006}. These, however, all compute diffraction and interference for an incoming and outgoing direction instantaneously at the location of reflection, whereas we defer the calculations of interference to a later stage. Zhang and Levoy were the first to introduce the connection between rays and the WDF in computer graphics~\shortcite{Zhang2009}. The Augmented light field (ALF), inspired by the WDF, was presented by Oh et al.~\shortcite{Oh2009,Oh09}, describing how transmission of light through a mask can be modeled using ray based rendering techniques. Optical Path Differencing (OPD) techniques keep track of the distance a ray travels and calculate its phase. Ziegler et al. developed a wave--based framework~\shortcite{Ziegler2008}, where complex values can be assigned for occluders to account for phase effects. They also implemented hologram rendering based on wave propagation (with the spatial frequency)~\shortcite{Ziegler2007}. Edge diffraction allows speedup by first searching for diffracting edges and then creating new sources at those positions~\cite{Freniere1999,tsingos2000}. In contrast, our WBSDF indirectly encodes the phase information in the reflected radiances and directions by introducing negative real coefficients. Therefore, no modification to the rendering framework is necessary.\\

\noindent\textbf{Wave--based Audio Rendering:} For efficient rendering of sound, many techniques assume that high frequency sound waves can be modeled as rays. The two main diffraction models used in these geometric simulations are the Uniform Theory of Diffraction (UTD)~\cite{Kouyoumjian1974,Tsingos2001} and the Biot--Tolstoy--Medwin (BTM) model~\cite{hothersall1991,Torres2001}. These techniques, which are often referred to as edge--diffraction, first search for diffracting edges and then create new sources at those positions. UTD based--methods scatter incoming sound beams in a cone around the edge, where the half--angle of the cone is determined by the angle that the ray hits the edge~\cite{Chandak2008,Rick2007}. They are often preferred over BTM because of their efficiency, but are less accurate in lower frequencies. BTM is more accurate and is also applicable for non--geometric acoustics~\cite{Torres2001}, but is not applicable for interactive renderings due to its complexity.\\

\noindent An overview and comparison of several diffraction and interference simulation methods is presented in Table~\ref{tb:overview_rendering_methods} and Section~\ref{sec:performance}.

\section{Formulation of a Wave Based--BSDF}
\label{sec:concept}
\begin{figure}[t!]
\centering
\includegraphics[width=\columnwidth]{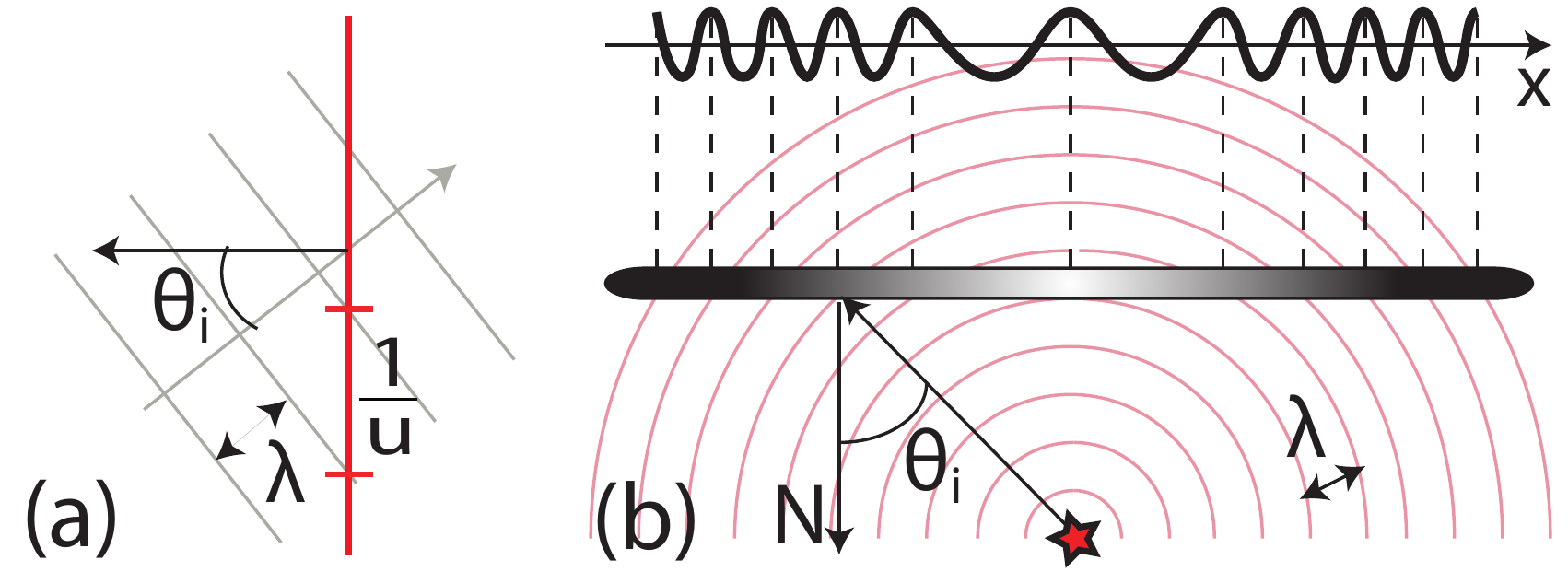}
\caption{Relationship between spatial frequency and incident angle. (a) Spatial frequency $u$ of incoming light is dependent on the incident angle $\theta_i$ and the wavelength $\lambda$ of the light. (b) The steeper incoming angle, the higher spatial frequency becomes.}
\label{fig:spat_freq}
\end{figure}



Consider the rendering equation~\cite{Kajiya1986}:
\begin{linenomath}
\begin{equation}
L_{o} ( x, \theta_{o}, \lambda ) = L_{e} ( x, \theta_o, \lambda ) + \int \rho ( x, \theta_o, \theta_i, \lambda ) L_i( x, \theta_i, \lambda ) \textrm{d} \theta_i, \label{eq:render_eq}
\end{equation}
\end{linenomath}
where $L_o( x, \theta_o, \lambda )$ is the total outgoing light from the point $x$ in the direction $\theta_o$ with wavelength $\lambda$. $L_e$ describes the light emitted at the point $x$ in the same direction, and $L_i$ denotes the incoming light at the point from a certain direction $\theta_i$. The function $\rho$ represents the proportion of scattered light for a given incoming and outgoing direction, and is often referred to as the Bidirectional Scattering Distribution Function (BSDF).

As Eq.~\ref{eq:render_eq} only takes the intensity of rays into account but not phase information, interference between multiple rays cannot be described unless travel distances of individual rays are tracked. Previous methods have described how to add local diffraction and interference effects to the Bidirectional Reflection Distribution Function (BRDF)~\cite{Stam1999}. However, a traveling ray does not carry phase information and therefore is unable to interfere in a later stage. It is challenging to represent this deferred interference in ray space. The Wigner Distribution Function, however, is a convenient method to understand and represent diffraction and interference.

The Wigner Distribution relates the spatial $x$ and spatial frequency $u$ content of a given function. While it can be applied to many different functions (i.e. music, images, etc.), it is a particularly useful description of a wave.
The Wigner Distribution of a 1D complex function of space $t(x)$ can be defined as
\begin{equation}
{W}_t(x,u) = \int { t\left( x + {x' \over 2} \right) t^{*}\left( x - {x' \over 2} \right) \textrm{e}^{-i 2 \pi x' u} \textrm{d}x'},\quad
\label{eq:R_WDF}
\end{equation}
where $^*$ is the conjugate operator. The function
\begin{equation}
J(x, x') = t\left( x + {x' \over 2} \right) t^{*}\left( x - {x' \over 2} \right)
\end{equation}
is often called the mutual intensity~\cite{Bastiaans2009} and is the correlation function of a complex--valued microstructure geometry $t(x)$. Note that after the Fourier transform of the mutual intensity, the WDF contains only real values, positive as well as negative, since the mutual intensity is Hermitian. In this section we use a 1D function as input to explain the concept of using the WDF, but the extension to a 2D input signal is straightforward as
\begin{equation}
{W}_t(x,y,u,v) = \int \int { J(x,y, x', y')  \textrm{e}^{-i 2 \pi (x' u + y' v)} \textrm{d}x' \textrm{d}y'},\quad
\end{equation}
where
\begin{equation}
J(x,y, x', y') = t\left( x + {x' \over 2}, y + {y' \over 2} \right) t^{*}\left( x - {x' \over 2}, y - {y' \over 2} \right).
\end{equation}

The essence of this representation is that a complex wavefront is decomposed into a series of plane waves with spatial frequency $u$ and a real-valued amplitude. This spatial frequency corresponds to the directionality of the parallel wavefront~\cite{Goodman2005} as illustrated in Figure~\ref{fig:spat_freq}(a-b).  The local spatial frequency is related to the wavelength and the direction of wave; which is normal to the wavefront, as
\begin{linenomath}
\begin{equation}
u = \frac{\sin \theta_{i} }{\lambda}.
\end{equation}
\end{linenomath}


The Wigner Distribution is often used for describing the complex--valued wavefront arising from surfaces like a grating. If a plane wave (i.e. light from a point light source at infinity) hits a surface, the outgoing wavefront function $R_t$ can be described with this Wigner Distribution Function using Eq.~\ref{eq:R_WDF} as
\begin{equation}
R_{t}(x,u) = W_{t}(x,u)  
\end{equation}



\begin{figure}[t!]
\centering
\includegraphics[width=\columnwidth]{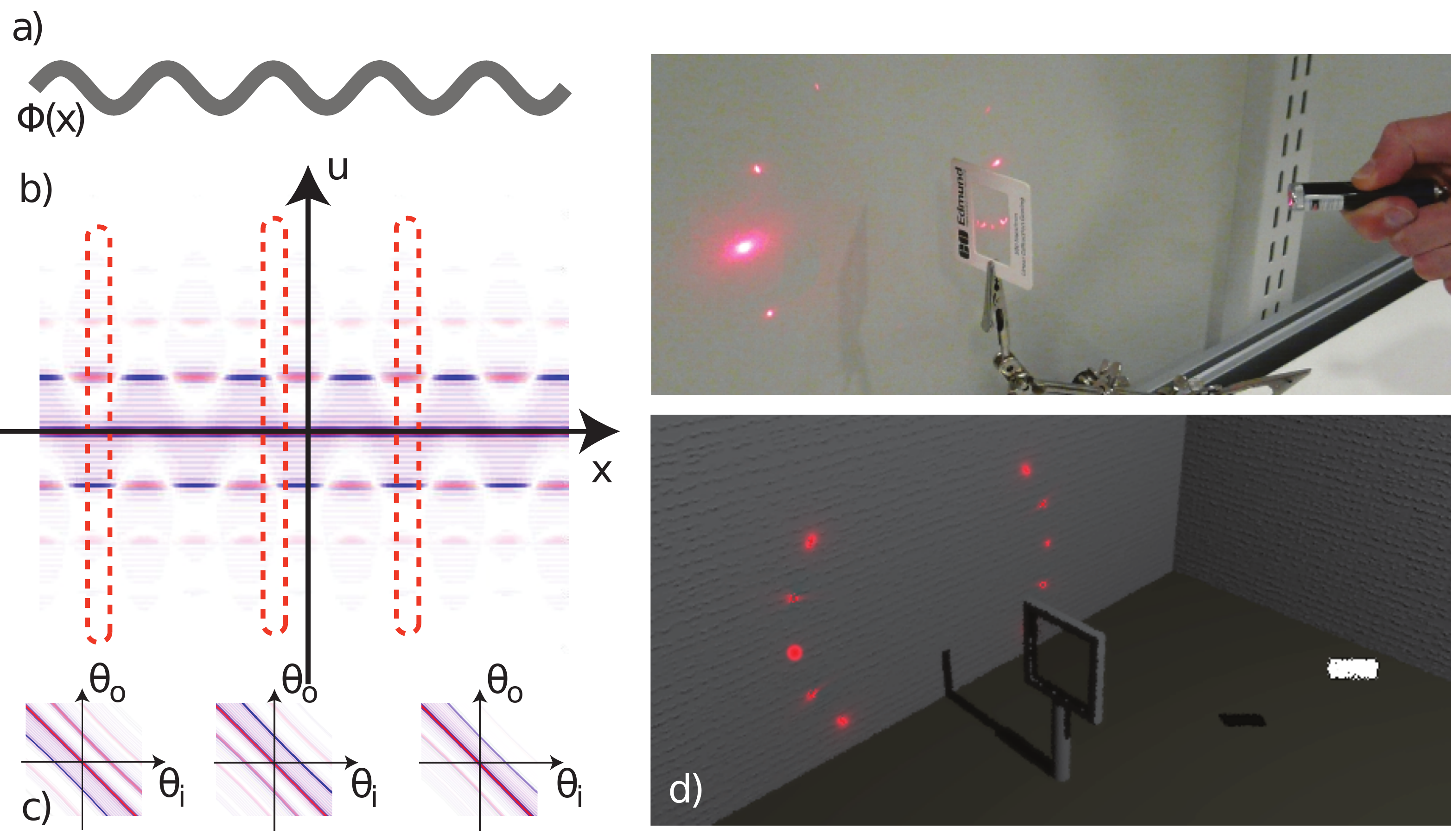}
\caption{Creating WBSDF from microstructure. (a) A simple sinusoidal grating is represented as a phase function. (b) Its WDF in position--spatial frequency $(x,u)$ domain, where red defines positive values and blue negative (c) We convert the vertical slices into plots for incoming and outgoing angle, creating the WBSDF (shown at three different locations on the microstructure) (d) Rendering using this WBSDF. Top-right shows a photo for visual validation.}
\label{fig:brdf}
\end{figure}

When a more complex wavefront hits the surface, we have to decompose the incoming wavefront $R_i$ into plane waves and transform each to reconstruct the outgoing wavefront. This transformation is similar to the rendering equation: 
\begin{linenomath}
\begin{equation}
R_o( x, u_o ) = \int {W}_t ( x, u_o, u_i ) R_i( x, u_i ) d u_i,
\end{equation}
\end{linenomath}
where $R_o$ is the transmitted wavefront for an incoming wavefront $R_i$. $W_t( x, u_o, u_i )$ is solely dependent on the microstructures of geometry $t(x)$. Under the assumption that the structure is sufficiently thin, this transformation becomes angle shift--invariant; the transmitted pattern simply rotates with the incident rays. Note that this property does not necessarily imply isotropy, as the transmitted wavefront does not have to be symmetric. An angle--shift invariance assumption is valid when the lateral size of a patch is significantly larger than the thickness of the microstructure. Under this assumption the transmission function is given by
\begin{linenomath}
\begin{equation}
R_o( x, u_o ) = \int {W}_t ( x, u_o - u_i ) R_i( x, u_i ) d u_i.
\label{eq:transmission}
\end{equation}
\end{linenomath}
This transformation is also calculated using the WDF as in Eq~\ref{eq:R_WDF}.
Note that $W_t(x,u)$ may contain positive as well as negative real--valued coefficients. However, after the integration over a finite neighborhood, the \textit{intensity} value always becomes non--negative~\cite{Bastiaans1997}. For reflections we assume that the reflected wavefront is a mirror of the transmitted wavefront, and therefore:
\begin{linenomath}
\begin{equation}
R_o( x, u_o ) = \int {W}_t ( x, - u_o - u_i ) R_i( x, u_i ) d u_i.
\label{eq:reflection}
\end{equation}
\end{linenomath}
This leads to the light transport defining \textit{wave rendering equation} for thin microstructures (compare with Eq. \ref{eq:render_eq}),
\begin{linenomath}
\begin{equation}
R_{o} ( x, u_o ) = R_{e} ( x, u_o ) +
 \int {W}_t ( x, - u_o - u_i ) R_i( x, u_i ) \textrm{d} u_i.
 \label{eq:render_eq_waves}
\end{equation}
\end{linenomath}

\subsection{Converting Microstructures to WBSDF}
\label{sec:convertBRDF}

In order to create the WBSDF, we need to know the exact microstructure $t(x)$ of the material. The microstructure is defined by an amplitude $a(x)$ and phase $\Phi(x)$ with
\begin{linenomath}
\begin{equation}
t(x) = a(x) \textrm{e}^{i \Phi(x)}.
\label{eq:microstructure}
\end{equation}
\end{linenomath}
Equation~\ref{eq:microstructure} allows us to calculate the WDF for the microstructure by Eq.~\ref{eq:R_WDF}.
Finally, the BSDF is calculated using a basis transformation as
\begin{linenomath}
\begin{equation}
\rho( x, \theta_i, \theta_o, \lambda ) \sim {W}\left( x, \frac{ \pm \sin \theta_o  - \sin \theta_i}{\lambda} \right),
\label{eq:brdf_as_wdf}
\end{equation}
\end{linenomath}
with positive sin($\theta_o$) in the case of transmission and a negative for reflection.\\
 
The amplitude $a(x)$ of the microstructure defines transmissivity, reflectivity and absorption. For example, an open aperture is a function that is $1$ within the opening and $0$ elsewhere. The component $ \Phi(x)$ indicates the phase delay introduced due the refraction index or the thickness of the material at position $x$. Figure~\ref{fig:brdf}(a) illustrates the phase function for a sinusoidal grating in 2D.\\

As an example, consider the sinusoidal grating in Figure~\ref{fig:brdf}. We can formulate the microsurface as
\begin{equation}
t(x) = \textrm{e}^{i \frac{m}{2} \textrm{sin}(2 \pi \frac{x}{p})}
       = \sum_{q=-\infty}^{\infty} J_q \left( \frac{m}{2} \right) \textrm{e}^{i 2 \pi q \frac{x}{p}}
\end{equation}
where $m$ is the maximum height of the grating, $p$ is the period and $J_q$ is the Bessel function~\cite{Goodman2005}. Using Eq.~\ref{eq:R_WDF} we can calculate its Wigner Distribution Function as
\begin{multline}
{W}(x,u) = \sum_{q_1=-\infty}^{\infty} \sum_{q_2=-\infty}^{\infty} J_{q_1} \left( \frac{m}{2} \right) J_{q_2} \left( \frac{m}{2} \right)\\
  \times \textrm{e}^{i 2 \pi \frac{x}{p} (q_1 - q_2)} \delta\left( u - \frac{q_1 + q_2}{2p}\right).\qquad
\end{multline}
As a final step we convert this to the Wave BSDF using Eq.~\ref{eq:brdf_as_wdf}.\\

\subsection{Creating a Statistical WBSDF}

\begin{figure}[t!]
\begin{center}
 \centerline{\includegraphics[width=0.8\columnwidth]{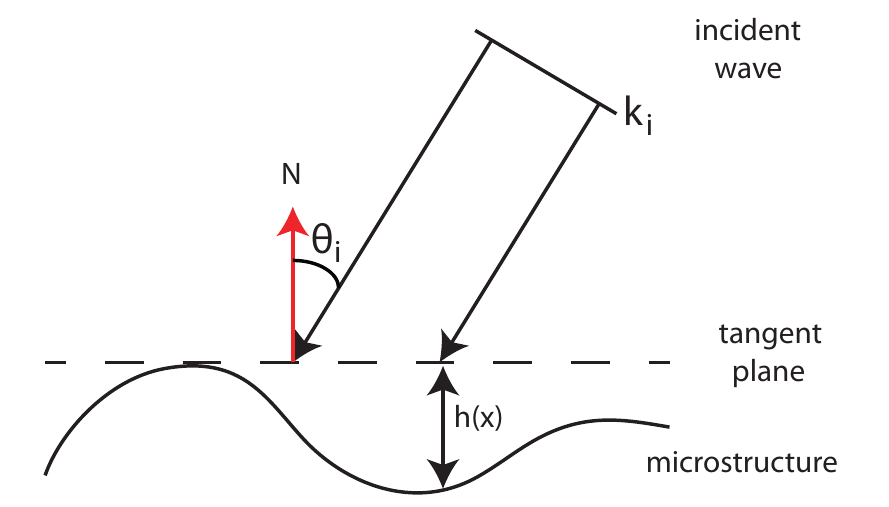}}
 \caption{A schematic overview of the approximated surface field, where $k_i$ is the center wave vector incident at the elevation angle $\theta_i$. $h(x)$ represents the height difference from the target plane to the actual surface.}
 \label{fig:statistical}
\end{center}
\end{figure}

We can easily express the microstructure for a simple surface. Eq.~\eqref{eq:R_WDF} implies that the BSDF can be computed provided the exact surface profile of the material. However, the exact microstructure may not be known for challenging objects. One way to circumvent this is to represent such surfaces statistically via autocorrelation and standard deviation (articulating periodicity and roughness, respectively).  In this situation, we can compute the WBSDF with a statistical average as
\begin{equation}
{W}_{t}(x,u) = \int \Big< t\left(x+\frac{x'}{2}\right) t^{*}\left( x-\frac{x'}{2}\right) \Big>\textrm{e}^{-i2\pi x' u}\text{d}x',
\end{equation}
where $\langle~\rangle$ denotes average. Depending on the surface properties and rendering environments, different types of statistical averaging can be used. In general, Gaussian statistics is assumed and statistical parameters such as standard deviation $\sigma$ or autocorrelation length can be tuned. Typically a standard deviation of height $\sigma_{h}$ is normalized by the wavelength $\lambda$ of light as
\begin{equation}
\sigma = \frac{\sigma_h}{\lambda}.
\end{equation}
This statistical average approach has been used in the diffraction shader~\cite{Stam1999} and BRDF estimators~\cite{Hoover2006}. Note that $\langle t\left(x+x'/2\right) t^{*}\left( x-x'/2\right) \rangle$ is mutual intensity $J(x, x')$. 
According to Goodman~\shortcite{Goodman1984}, who calculated the statistical properties of laser speckles based on the tangent--plane approximation, the surface field can be expressed as 
\begin{eqnarray}
\displaystyle
t(x) &=& a(x) \textrm{e}^{i k_i x} \textrm{e}^{\frac{2 \pi i}{\lambda} (1 + \textrm{cos} \theta_i ) h(x)}\\
       & = & a(x) \textrm{e}^{i k_i x} \textrm{e}^{i \alpha(x)} .
\label{eq:surface}
\end{eqnarray}
where $k_i$ is the center wave vector incident at the elevation angle $\theta_i$ and $h(x)$ is the height difference between the tangent--plane and the actual surface as illustrated in Figure~\ref{fig:statistical}. We assume no shadowing and no multiple interreflections. Also, the roughness should be $\sigma \gg 1$.

From Eq.~\eqref{eq:surface} we can derive the phase variance
\begin{equation}
\sigma_{\alpha}^2 = \left[ 2 \pi \sigma ( 1 + \cos \theta_i ) \right]^2
\end{equation}
and the phase autocorrelation
\begin{equation}
R_\alpha( x' ) = \left[ \frac{2 \pi}{\lambda} ( 1 + \cos \theta_i ) \right]^2 R_h(x'),
\end{equation}
where $R_h(x')$ is the autocorrelation of the surface height (Figure \ref{fig:statistical-results}(a)).
If we assume a Gaussian distribution of surface height we can now formulate the correlation function as
\begin{eqnarray}
\displaystyle
\gamma( x' ) &=& \textrm{e}^{i k_i x'} \langle \textrm{e}^{i \alpha(x) - \alpha( x - x')} \rangle\\
                     &=& \textrm{e}^{i k_i x'} \textrm{e}^{-\sigma_\alpha^2 \left[ 1 - \rho_h( x' ) \right]},
\end{eqnarray}
where 
\begin{equation}
\rho_h(x') = \frac{R_h(x')}{\sigma_h^2}.
\end{equation}

Since the WDF is defined with respect to the correlation function as in Eq.~\eqref{eq:R_WDF}, we can derive the WBSDF from the WDF of the correlation function as 
\begin{linenomath}
\begin{multline}
\mathcal{W}_{\gamma}(u) = \int J(x') \textrm{e}^{-i 2\pi u \cdot x'}\text{d}x'\\
 = \textrm{e}^{-\sigma_{\alpha}^{2}} 
\mathcal{F} \left[\exp\left\{\frac{\sigma_{\alpha}^{2}}{\sigma_{h}^{2}}R_{h}(x')\right\} \right]\Bigg|_{x' \rightarrow u - \frac{\sin\theta_i}{\lambda}}.
\label{eq:WDF_gamma}
\end{multline}
\end{linenomath}


 \subsection{Benefits}
\textbf{Near and far field: }
The WDF as well as the WBSDF are conserved along rays in the paraxial region~\cite{Bastiaans2009}, hence they are valid in both the near-field and far-field. In the far-field, the observed wave at a single point is only dependent on angle, and is essentially independent of the distance from the grating. Near-field corresponds to the Fresnel region in optics (not to be confused with the near-zone in optics where the evanescent field is still strong). The near-field is the region close to the grating, where the wave's distance to the grating also influences the observed pattern. In contrast to previous diffractive BRDFs~\cite{Stam1999}, the WBSDF does not require an assumption that the object and receiver are at infinity.
Near field effects, such as lenses, require more computation than far field effects. Hence, importance sampling is highly desired.\\

\noindent\textbf{Importance sampling and global illumination: } The WBSDF uses instantaneous diffraction; i.e., the magnitude of energy scattered in all directions is known at the surface. This allows importance sampling for an efficient light simulation and is also suitable for rendering global illumination.\\

\noindent\textbf{Light coherence: } Natural scenes are composed of incoherent sources. This means that each single point source (at a reasonable distance) is coherent with itself and mutually incoherent with other point sources. Simulating interference effects normally requires rendering the scene for each light source independently, and then summing up images to create the final result. Because creating diffractive effects for a single light source only consists of summing up photons (positive or negative), our technique can simultaneously render coherent light as well as multiple incoherent light sources, such as area lights. In rare scenarios, multiple sources are mutually coherent. Here, we treat all the sources jointly to compute a single emissive WBSDF.

\subsection{Limitations}
We use the paraxial approximation, where incoming and scattered (diffracted) light propagate not far from the optical axis. To improve the accuracy in the non--paraxial zone, one can use the angle--impact WDF~\cite{Alonso2004}. Regarding polarization, we only consider linearly polarized light. Adopting the coherency matrix~\cite{Tannenbaum1994}, we can extend our method beyond linearly polarized light. The current model of the WDF only encodes spatially varying phase and does not take temporal phase into account. To model transient responses, one can incorporate the spatial phase from the WDF with temporal frequency. 

Using the WBSDF, a single reflected ray off a surface may contain negative radiance, which is physically impossible. However, by integrating over a finite neighbourhood, for example in a camera aperture, the total amount of received light always becomes non--negative due to the properties of the WDF~\cite{Bastiaans2009}. Therefore, our model does not support pinhole camera models for direct lighting. More details on this topic are presented in the supplemental material.

\section{Wave--Based BRDF in Practice}
We show representative ray--based wave effect rendering as examples for global illumination and sound propagation.

\subsection{Light Rendering}

\begin{figure}[t]
     \centering
     \subfigure
     {
         \includegraphics[width=.225\textwidth]{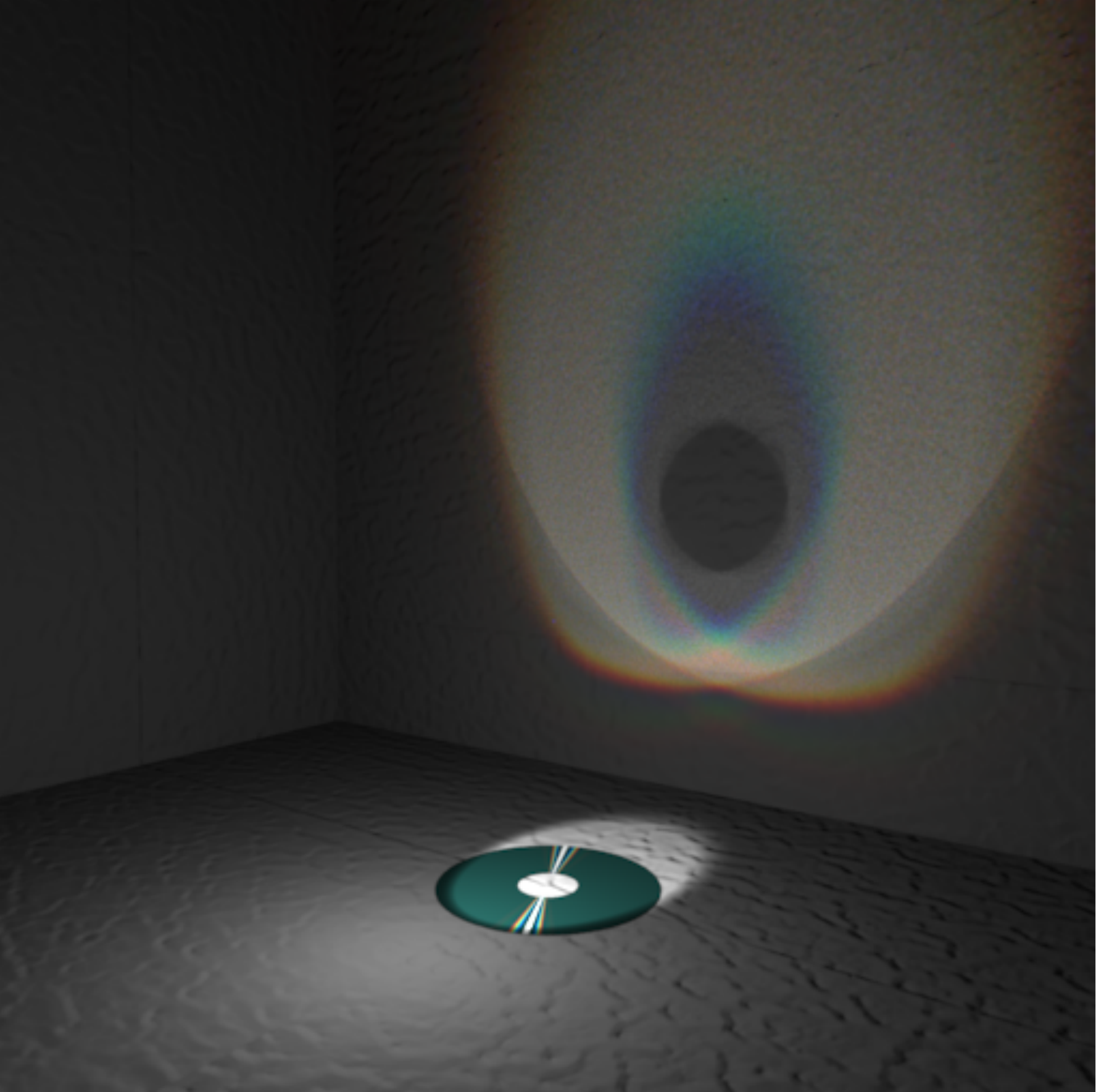}
     }	
     \subfigure
     {
          \includegraphics[width=.225\textwidth]{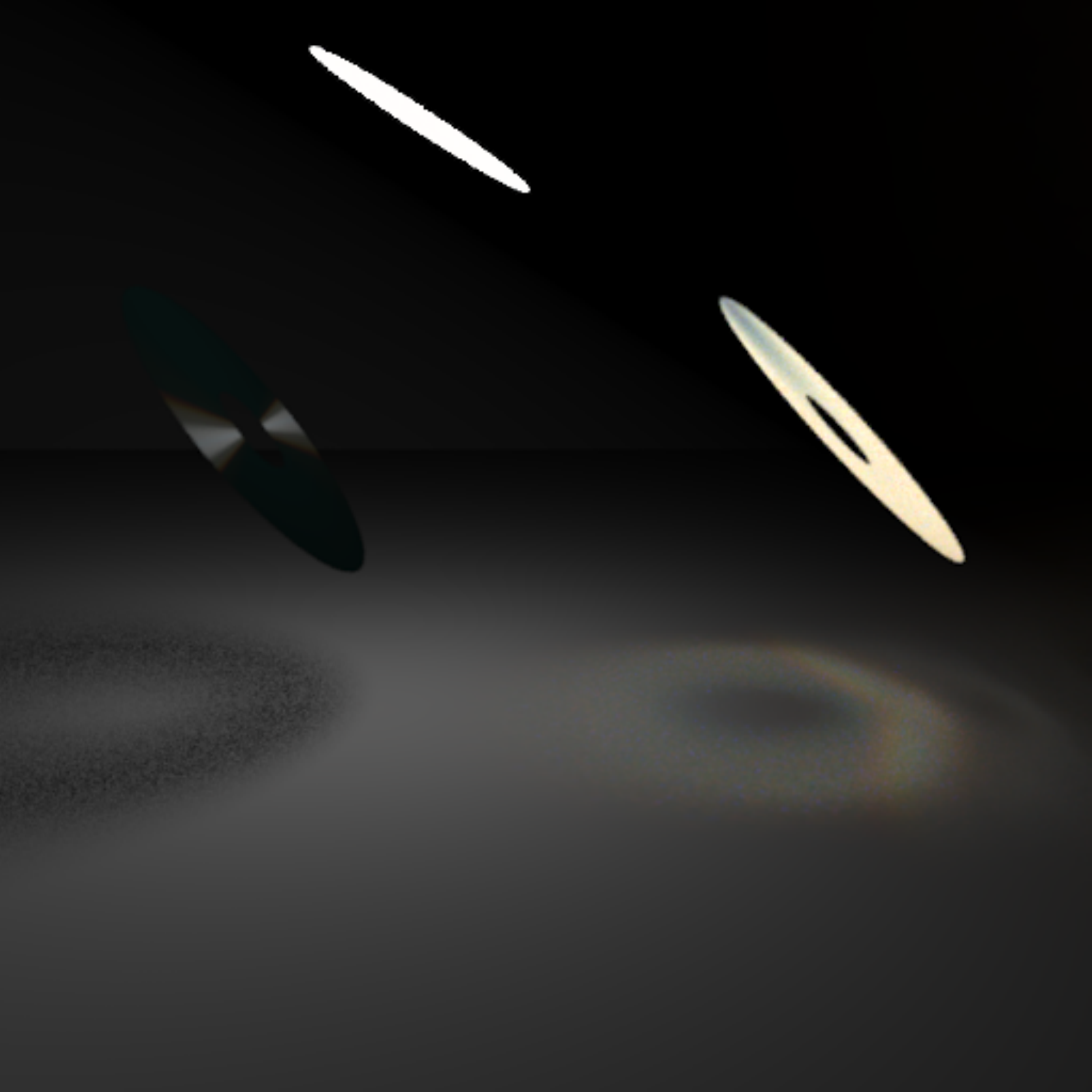}
     }	
     \caption{Renderings using the WBSDF under global illumination. (Left) Light reflected off a CD on a diffuse wall, (Right) light emitted from an area light (top) is reflected off the left CD onto the right CD and then to the floor.}
     \label{fig:results}
\end{figure}

\textbf{Diffraction and interference: } The WBSDF indirectly retains the phase information after diffraction, and defers the interference computation. This strategy supports interference after global illumination naturally. Figure \ref{fig:teaser} shows the diffracted wavefront from a CD undergoing refraction. As we describe later, instantaneous diffraction enables importance sampling. 

Figure~\ref{fig:results} shows 2 CDs interacting with each other and illuminating a wall. We model the CD as a phase grating with a pitch of $2~\mu  m$. We believe this is the first practical demonstration of global illumination of wave phenomenon, which is achieved using an unmodified PBRT.

\begin{figure}[h!]
\begin{center}
 \centerline{\includegraphics[width=1\columnwidth]{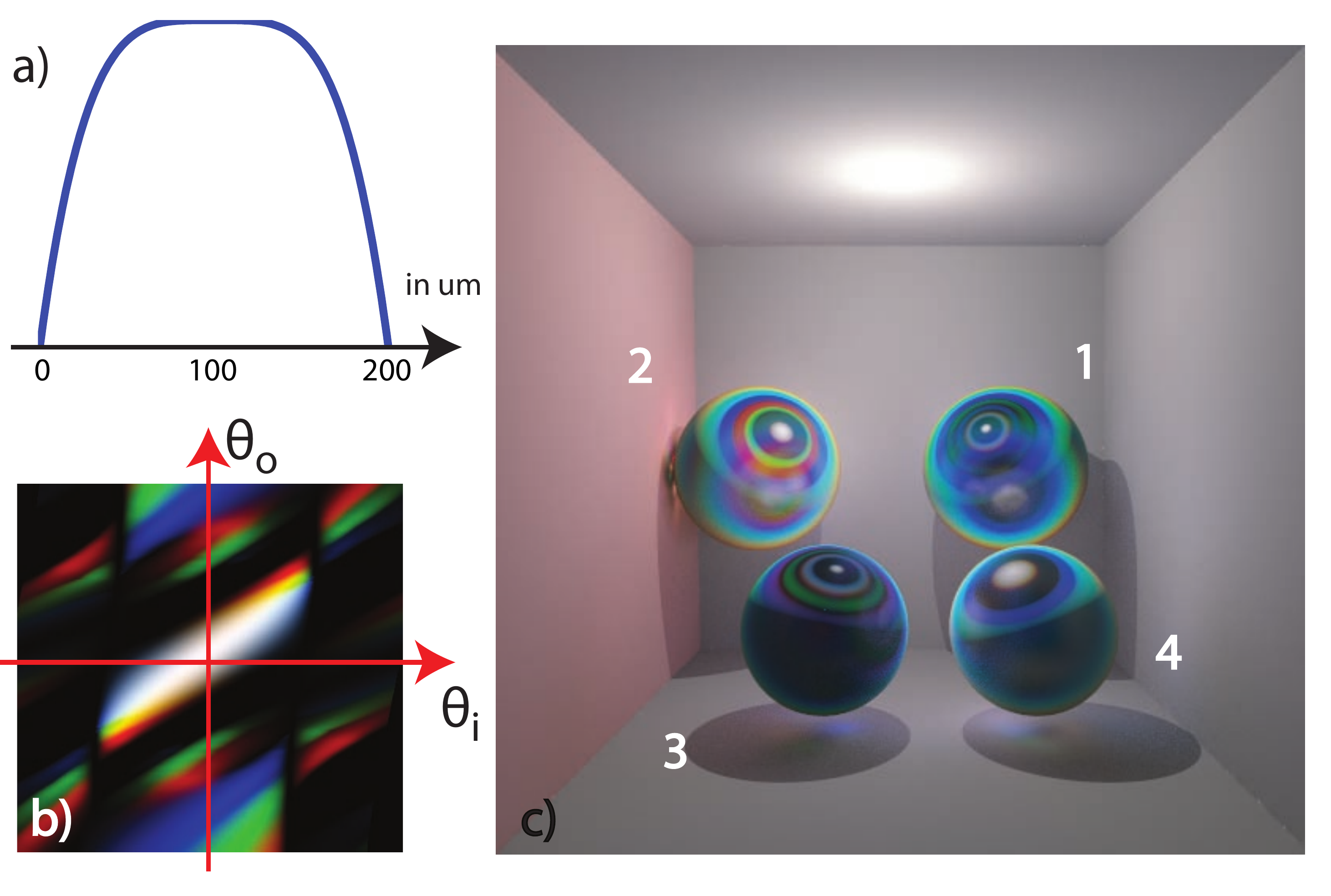}}
 \caption{Rendered results using the statistical approach of the WBSDF. (a) Example $R_{h}$ function describing the basic surface height of elements to reconstruct the microstructure. (b) The statistical WBSDF of element (a) plotted in $\theta_i$ and $\theta_o$ for a specific standard deviation. (c) Rendered result of spheres on which this BRDF is mapped. Spheres 1 to 4 have respectively a standard deviation of 10, 8, 6 and 4.}
 \label{fig:statistical-results}
\end{center}
\end{figure}

Figure~\ref{fig:statistical-results} demostrates the statistical approach of the BSDF. In the autocorrelation  of the surface we used the function 
\begin{equation}
R_h = -(x/a)^4,
\end{equation}
as shown in Figure~\ref{fig:statistical-results}(a). The BSDF is calculated for $\theta_{in}$ and $\theta_{out}$ (Figure~\ref{fig:statistical-results}(b)) and mapped onto spheres, each with a different standard deviation. 

\begin{figure}[b]
\centering
\includegraphics[width=.45\textwidth]{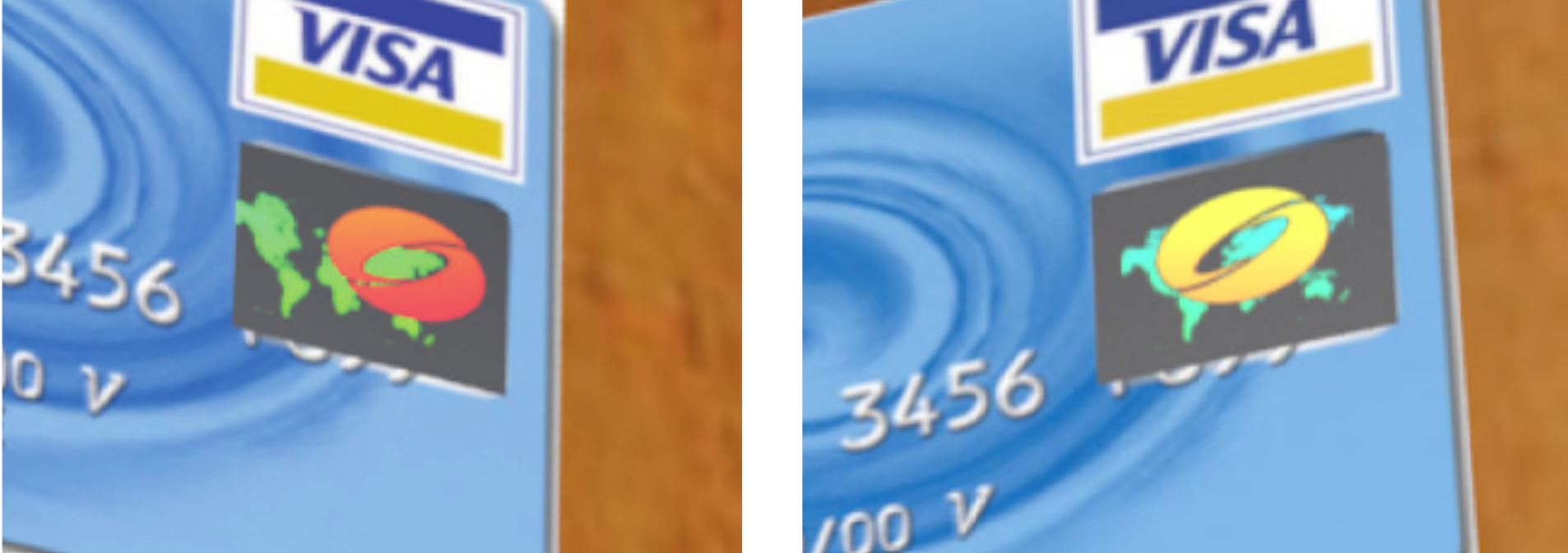}
\caption{Rainbow hologram renderings from different viewpoints. Instantaneous diffraction using the WBSDF pre-computes the view--dependent appearance.}
\label{fig:holo_sketch}
\end{figure}

\noindent\textbf{Rainbow holograms: } As we can create the WBSDF of any diffractive optical element based on its microstructure, we can derive a WBSDF for a holographic surface as well. As diffraction is computed instantaneously for the entire surface, we do not keep track of travel distance from a given point to the hologram surface, as OPD does. The hologram can be rendered directly from the WBSDF and creates interference inside the aperture of the camera lens. Note that there are many different ways of computing the WDF of the object~\cite{Plesniak2005}. For simplicity, here we encoded incoherent objects in the hologram; however any arbitrary wavefront information can be stored in the WBSDF.

%

\subsection{Simulating Camera Optics}

In camera optics, aberrations as well as diffraction affect the Point Spread Function (PSF) of the system. Hence,  conventional optics design software suites provide analytical tools for aberrations and diffraction. However, they treat the two separately; i.e., ray tracing uses spot diagrams as an estimate of the PSF, whereas diffraction models produce PSFs directly from Fourier optics analysis. Note that a Fourier optics analysis typically assumes a shift invariant PSF, hence the diffraction for an off--axis PSF is assumed to be identical to the on--axis PSF, which is not true in practice. 

\begin{figure}[h!]
  \includegraphics[width=\columnwidth]{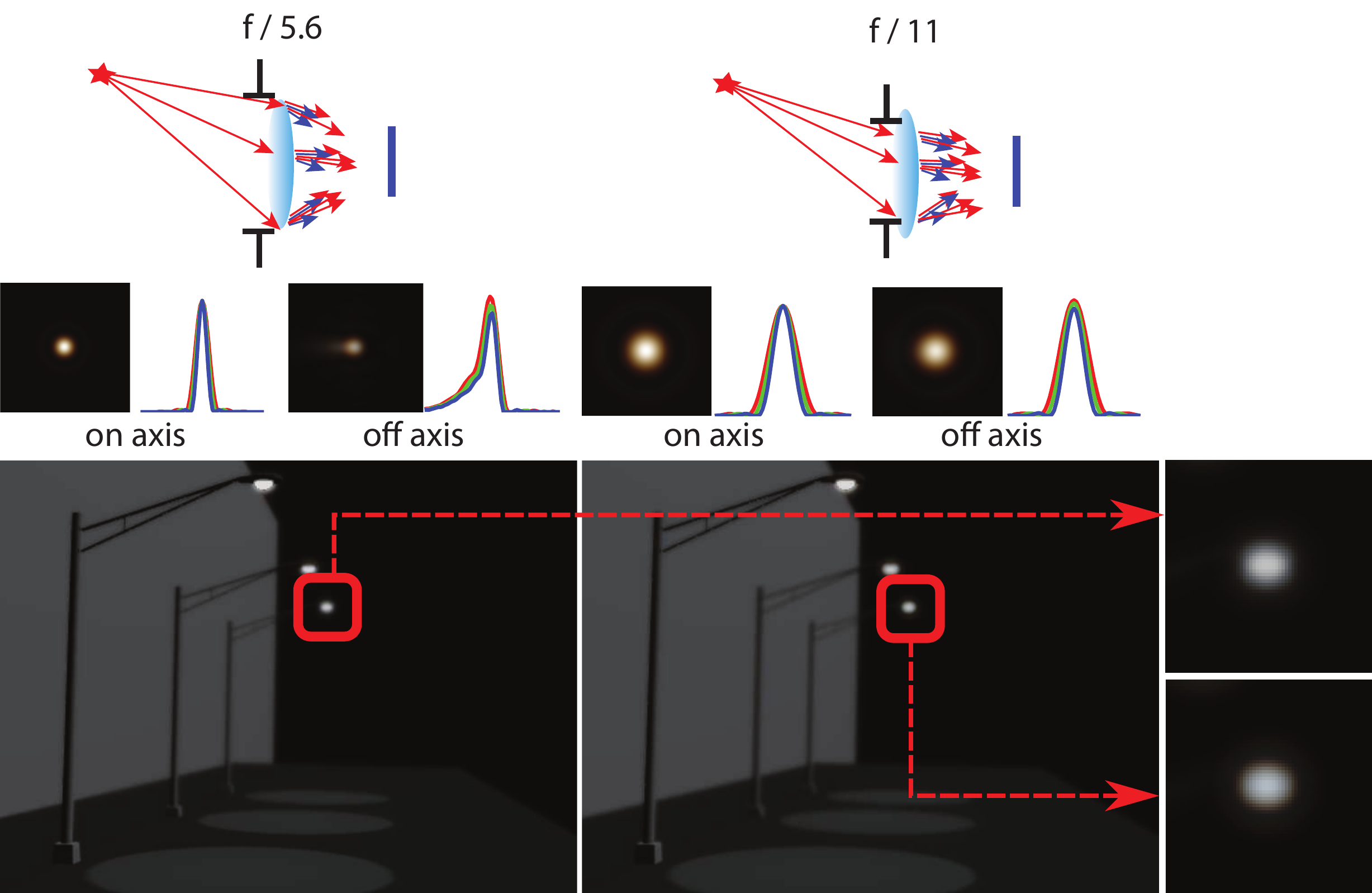}

\caption{Near field effects in camera lenses show the effect of an aperture creating diffraction and chromatic dispersion. (Left) PSFs of F/5.6 lens for lights at two different depths. (Right) PSFs of F/11 lens for two different depths.}
  \vspace*{-0.3cm}
  \label{fig:virt-camera}
\end{figure}

In addition to Oh et. al.~\shortcite{Oh2009}, our approach can simulate both aberrations and diffraction simultaneously for any location of the image plane, since it is based on ray--representation and able to include diffraction. We derived the WBSDF from the geometric structure of a camera lens with a circular aperture and calculated the PSF. Figure~\ref{fig:virt-camera} shows spatially varying PSFs for different lenses and point light positions. We then applied the PSFs to a computer generated image by convolution, simulating the appearance of the scene when viewed with this lens. Note that in this example, no chromatic aberration was assumed. Hence, the color dispersion results solely from diffraction. Although, we only consider thin lenses in this particular example, we can easily simulate a series of thick lenses as well. The global illumination ray--tracing takes care of refraction at air--glass interfaces and diffraction by an aperture is included by the WBSDF. 

\section{Implementation}

We achieve global illumination effects using an unmodified PBRT framework ~\cite{PBRT04}. The only minor change is that we comment out the check for non-negative radiance. To generate specific diffractive material, we simply created a new material plugin for PBRT. 

\begin{figure}[t!]
\centering
\captionsetup{type=table}
\includegraphics[width=\columnwidth]{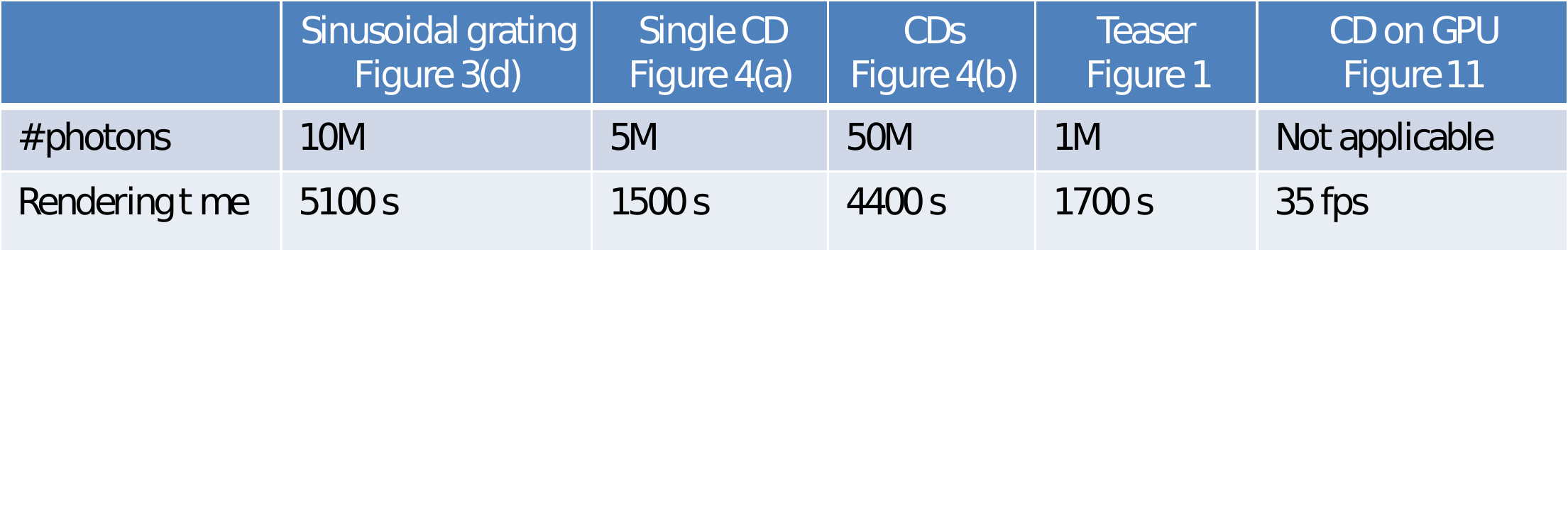}
\vspace{-2cm}
\caption{Rendering time and number of photons used for each scene. Implementation is done using PBRT. All these scenes are rendered single threaded on a 3GHz core. The GPU example was done on a NVidia 8400M with 640$\times$480 resolution.}
\label{tab:timings}
\end{figure}

We used photon mapping for Monte Carlo simulation of global illumination. Due to the nature of this technique, photon mapping performs well with our BSDF. Photons can convey negative energy to achieve destructive interference. WDF formulation ensures that non--negative final energy is gathered at a single point. This non--negativeness also applies to the gathering of photons in the photonmap.

\begin{figure}[h!]
\centering
\subfigure[Without importance sampling]{
\includegraphics[width=0.45\columnwidth]{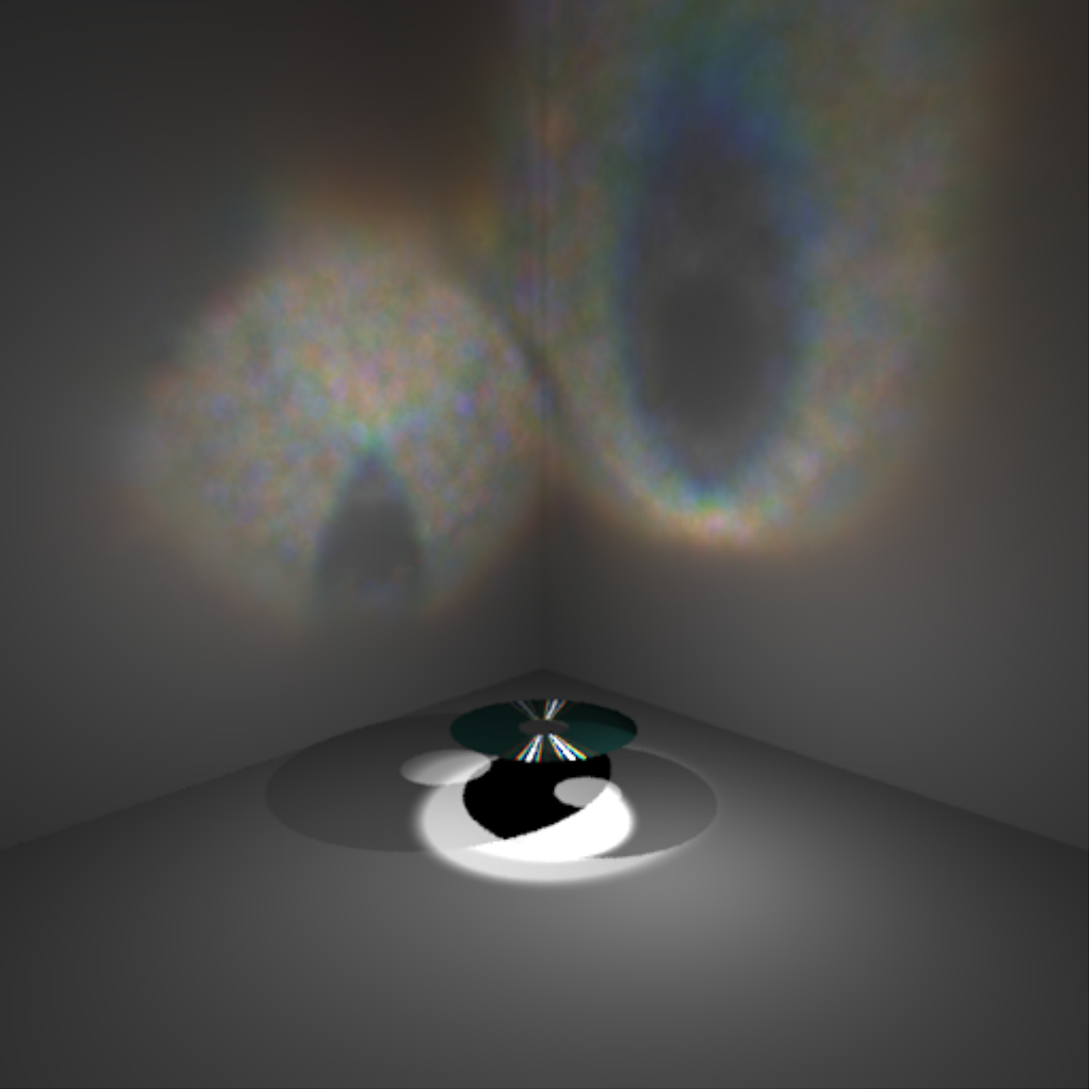}
}
\subfigure[With importance sampling]{
\includegraphics[width=0.45\columnwidth]{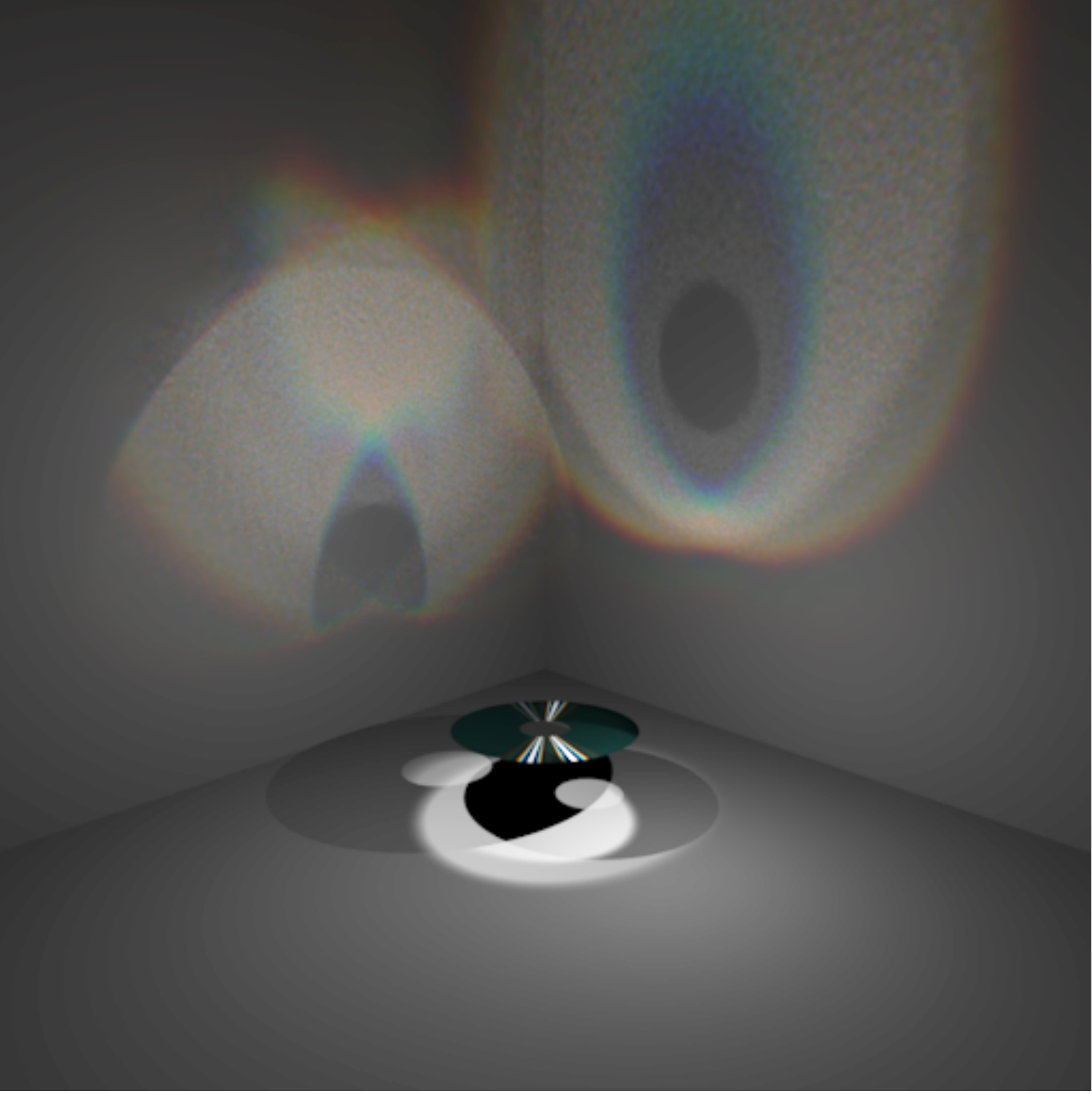}
}
\caption{Benefits of importance sampling (a) Without importance sampling, even after 14,000 s result does not converge (b) With, 1,300 s.}
\label{fig:importance_sampling}
\end{figure}

\begin{figure}[t!]
\centering
\subfigure{
\includegraphics[width=0.30\columnwidth]{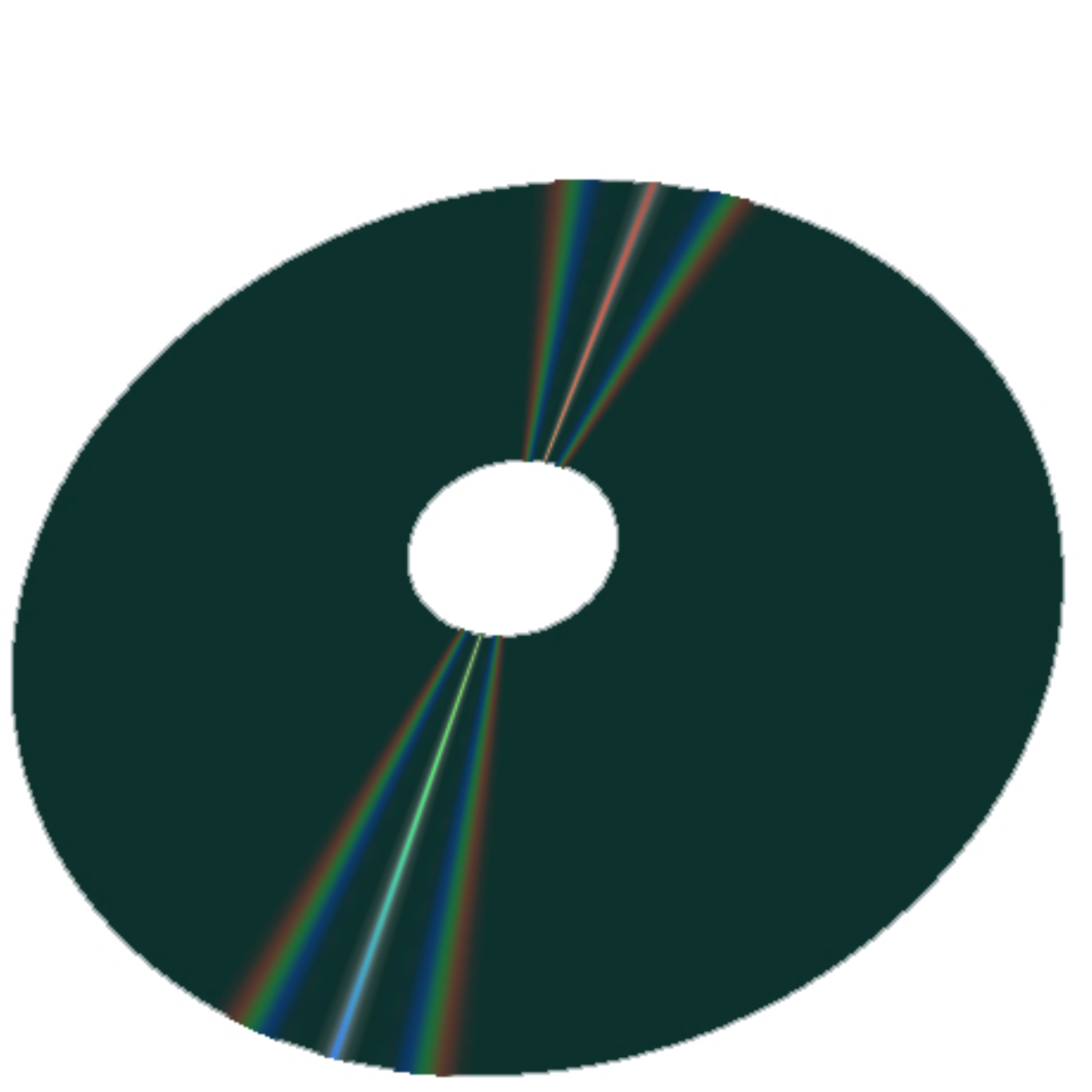}
}
\subfigure{
\includegraphics[width=0.30\columnwidth]{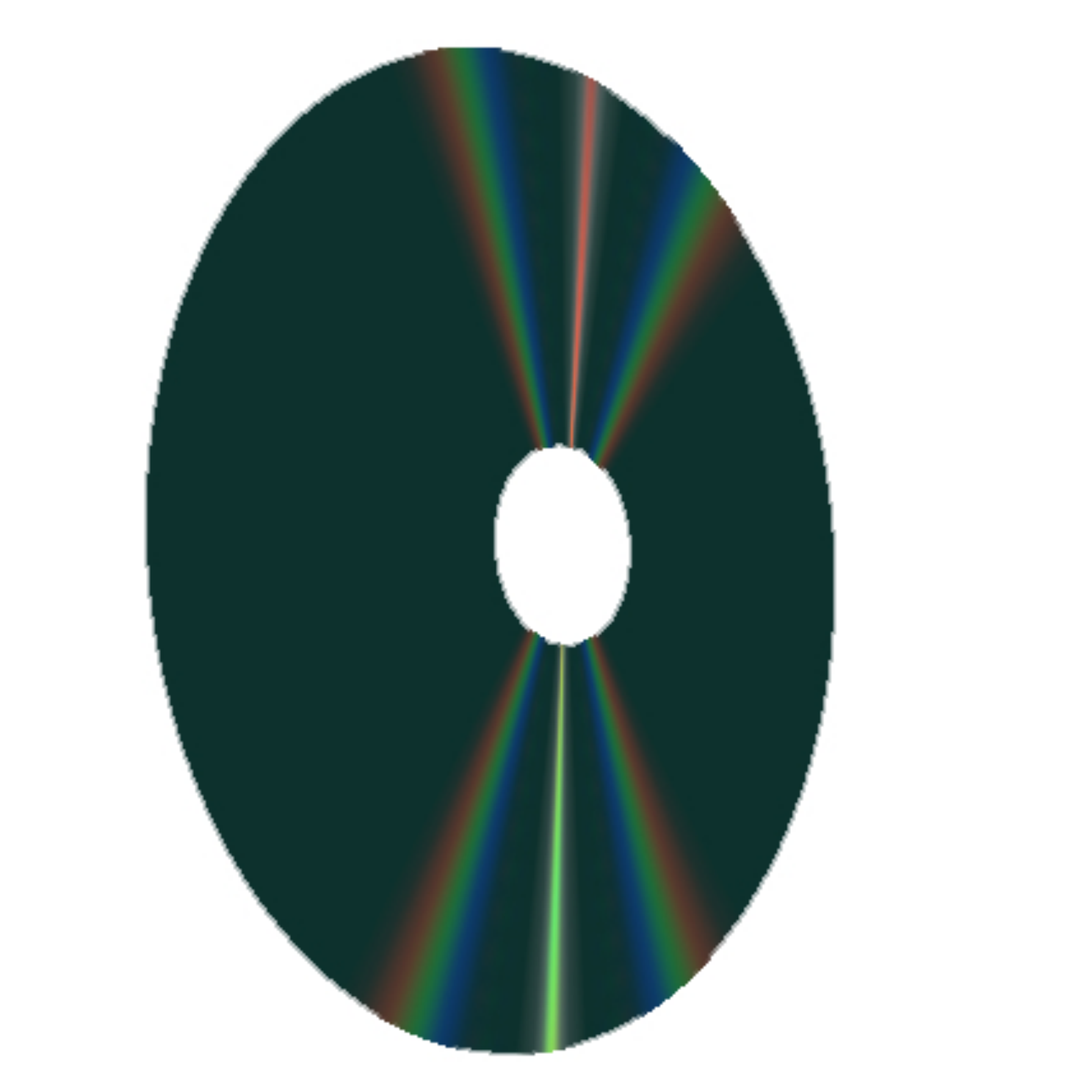}
}
\subfigure{
\includegraphics[width=0.30\columnwidth]{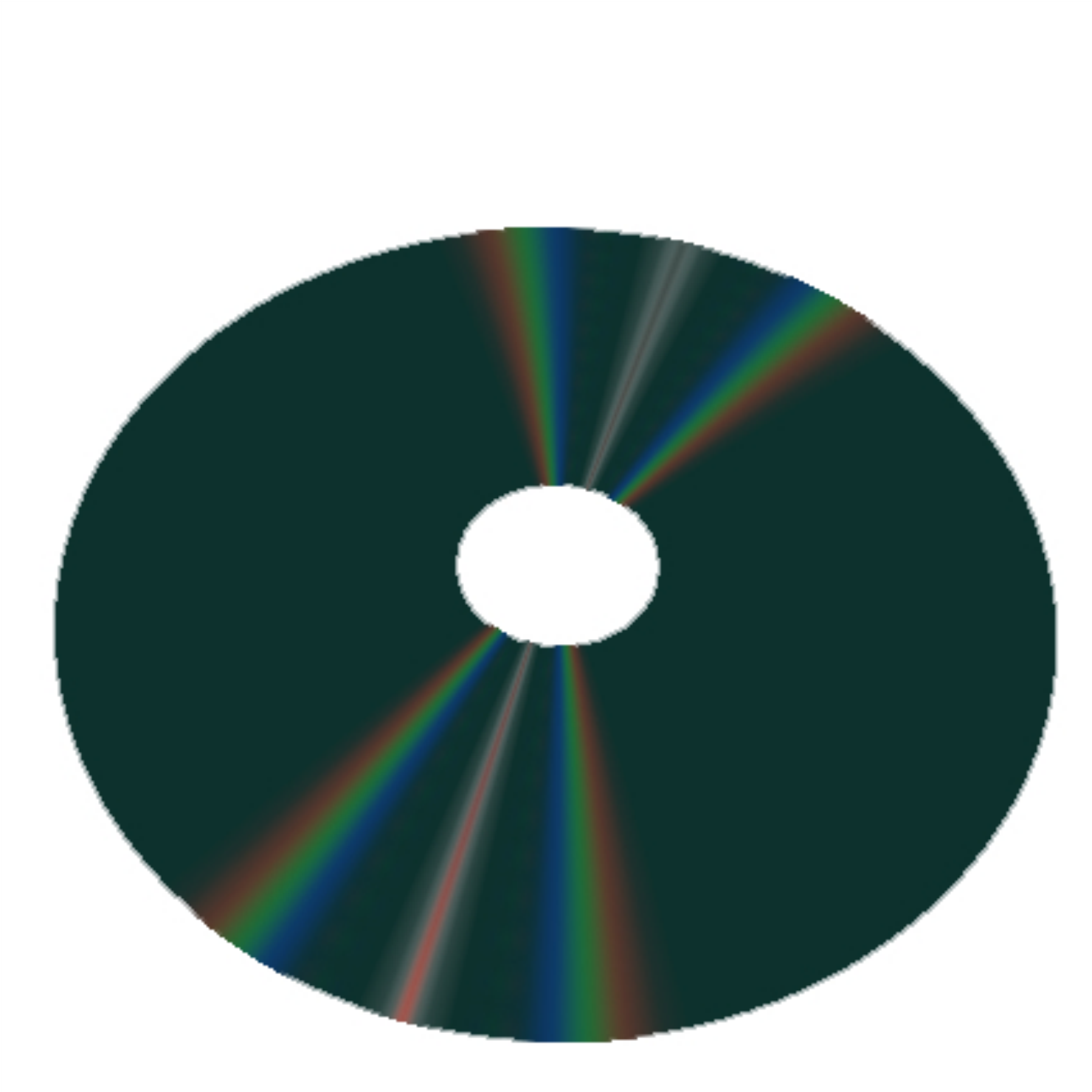}
}
\caption{CD rendered using the WBSDF in real-time on a GPU using look-up tables and mipmapping.}
\label{fig:gpu_cd}
\end{figure}

For materials with a separable WBSDF for $x$ and $y$ coordinates, we precomputed the WBSDF and saved it as a 2D look--up table, which allows fast calculation, importance sampling (Figure~\ref{fig:importance_sampling}), and real-time rendering of single bounce diffraction on the GPU (Figure~\ref{fig:gpu_cd}). As diffration has to be calculated on arrival, we need to sample over different locations for each pixel and integrate the values. Mipmapping calculates these interference effects efficiently. Similar to the diffraction shader implementation~\cite{Stam1999}, the colorfull 0th order highlight is added with a simple anisotropic shader~\cite{Ward1992}. Table~\ref{tab:timings} gives an overview of rendertime used for generating the images and audio examples. The pdf is calculated using the absolute values of the intensity in the BRDF lookup table.

Eq.~\eqref{eq:brdf_as_wdf} provides a function in position, angle, and wavelength. For structured surfaces, like sinusoidal gratings, we have a closed form solution to find which wavelengths have non--zero intensities. For computational efficiency, we tabulated the WBSDFs in terms of position and angle, and reduced the wavelengths to RGB values according to the camera response curves. The lookup tables, used for creating the examples in this paper, have a sampling rate of $0.02^\circ$ in angular domain and $48~\textrm{nm}$ in the spatial domain. For non--structured surfaces such as rainbow holograms, we computed Eq.~\eqref{eq:brdf_as_wdf} over 30 wavelengths and stored only RGB values using the appropriate weights.\\

\begin{figure}[b!]
\centering
\includegraphics[width=\columnwidth]{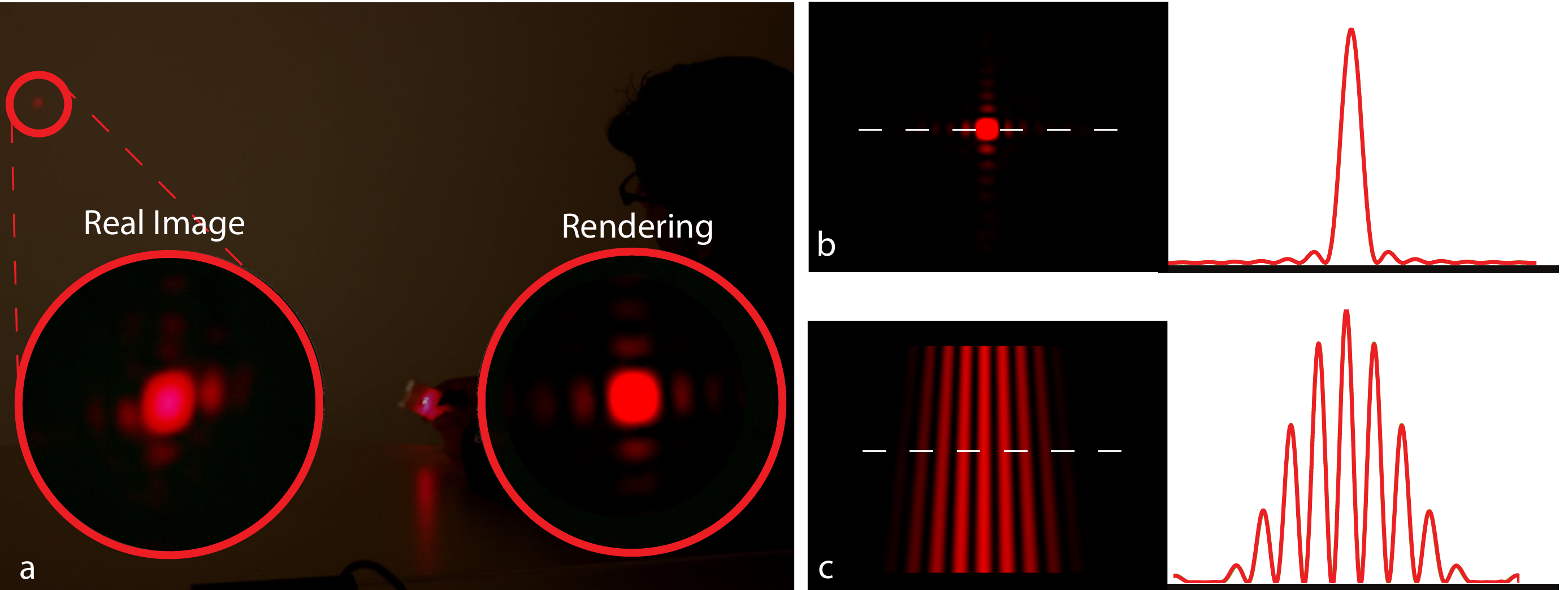}
\caption{Validation of the WBSDF (a) Visual comparison of photo and rendering of diffraction due to a laser through a rectangular aperture.  (b) Rendering of (a) and its 1D plot. (c) Two slit experiment and 1D plot. The 1D plots are compared with the Fraunhofer diffraction and have an absolute intensity error of less than 1\%.} 
\label{fig:accuracy}
\end{figure}

\noindent\textbf{Validation} We compared two representative diffraction examples with the one computed by Fourier optics and verified the accuracy of our technique. We used photon mapping to simulate Young's double slit experiment and light diffraction from a rectangular aperture (Fraunhofer diffraction). The absolute intensity error, mainly due to limited numerical precision, between the Fraunhofer diffraction pattern and the rendering with the WBSDF is less than 1\%. These results are presented in Figure~\ref{fig:accuracy}.

\section{Comparison}
\label{sec:performance}
\subsection{Comparison with OPD}
Even though optical path difference (OPD) provides accurate results of interference, it requires significant modifications to traditional rendering systems and is not able to perform importance sampling. Traditional raytracers commonly do not take distance or phase into account, and in order to implement OPD we need to make changes to the framework. The precision required for path length (for each wavelength) becomes challenging. In addition, OPD cannot exploit importance sampling. Determining the propagation direction of rays with dominant intensity in the presence of diffraction is challenging. OPD-based techniques must uniformly sample all the outgoing directions. Consider the example of the sinusoidal gratings of Figure~\ref{fig:brdf}. The WBSDF reveals the few important directions using our instantaneous diffraction theory. In more complex, global illumination settings, WBSDF-based approach is faster compared to OPD by the same factor as shown in Figure~\ref{fig:importance_sampling}. 

\subsection{Comparison with Diffraction Shaders}

\begin{figure}[t!]
\centering
\includegraphics[width=1\columnwidth]{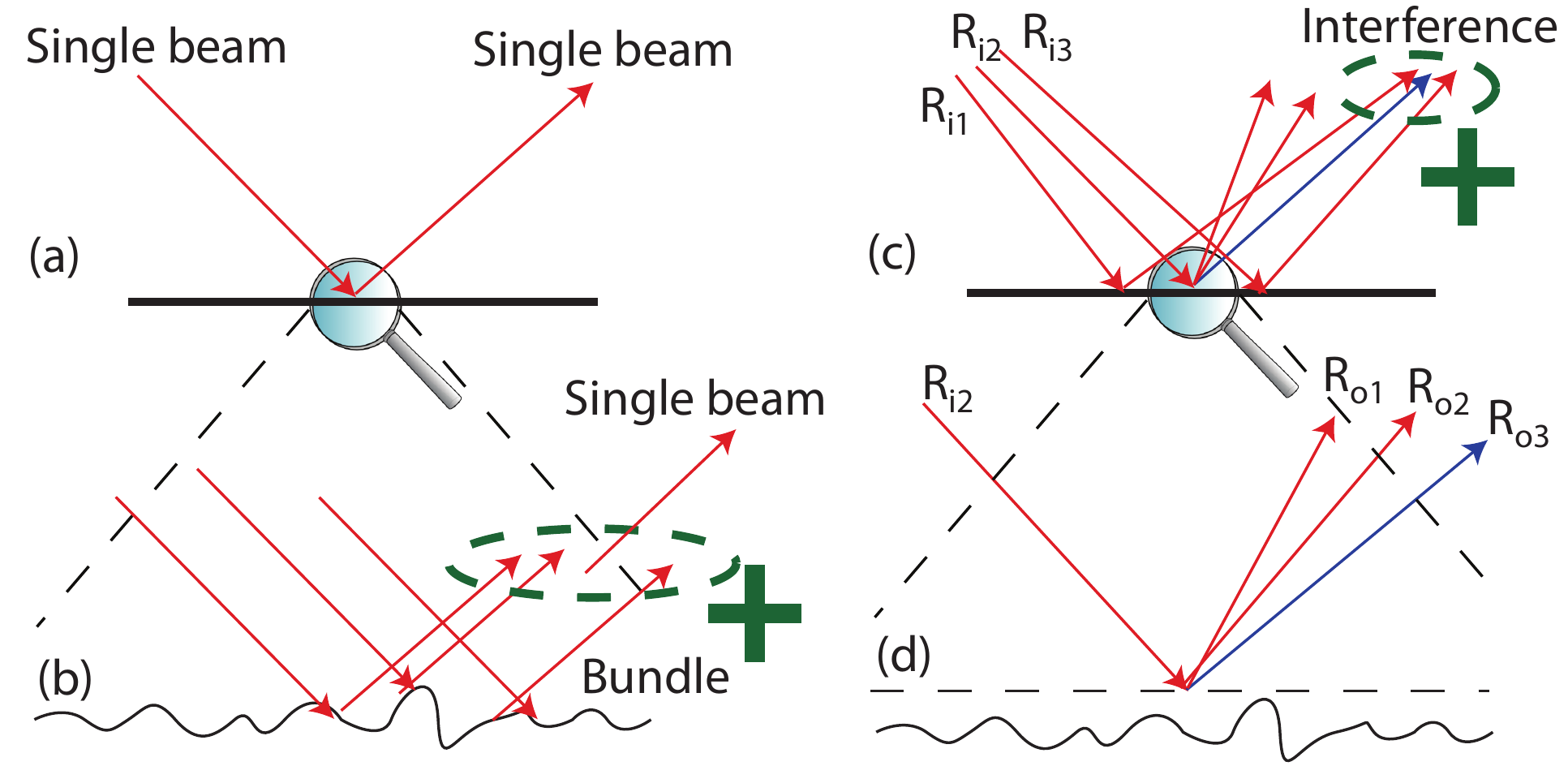}
\caption{Diffraction shaders is a special case of the WBSDF. (a--b) It pre-encodes the result of far-field interference into energy of the outgoing ray. These rays cannot destructively interfere anymore in a later stage. (c--d) WBSDF indirectly encodes phase information into the outgoing ray by introducing potentially negative radiance, allowing the rays to interfere later for global illumination.}
\label{fig:diff_shaders}
\end{figure}

Diffraction shaders (DS) pioneered the idea of fast and practical rendering of the far field diffraction of a single bounce ~\cite{Stam1999}. It turns out that the DS approach is a special case of our method where the light source and the observer are at infinity. DS converts parallel ray bundles predicted by the WBSDF and treats them as a single ray beam (Figure~\ref{fig:diff_shaders}). The WBSDF maintains a higher resolution representation and hence we can apply different integration kernels, transforms for propagation, and scattering models. The WBSDF also provides flexibility depending on incident/outgoing rays, camera geometry, computational speed and tolerance of error. The WBSDF supports a statistical model similar to the one used in DS for phase variations. 
Based on the assumption that  the light source and the observer are at infinity, we use the same notation as in Ref.~\cite{Stam1999} and present the relationship with the WBSDF as follows:
\begin{linenomath}
\begin{multline}
I\left(ku'\right) = \psi\left(ku'\right)\psi^{*}\left(ku'\right)\\
~~~\quad=\int \textrm{e}^{ikwh(x)}\textrm{e}^{iku'x}\textrm{d}x \int \textrm{e}^{-ikwh(x')}\textrm{e}^{-iku'x'}\textrm{d}x'\\
=\iint \textrm{e}^{ikwh(p+\frac{q}{2})}\textrm{e}^{-ikwh(p-\frac{q}{2})}\textrm{e}^{iku'q}\textrm{d}p\textrm{d}q~~~\\
=\int {W}_{d}(x, u') \text{d}x,\qquad\qquad\qquad\qquad\qquad\quad~~
\label{eq:diff_shader}
\end{multline}
\end{linenomath}
where $I$ is intensity, $\psi$ is field, $k=2\pi/\lambda$, $u' = \sin\theta_1-\sin\theta_2$, $w=-\cos\theta_1-\cos\theta_2$, $p=(x+x')/2$, $q=x-x'$, and ${W}_{d}(x,u')$ is the WDF of $\text{e}^{ikwh(x)}$ with respect to $\sin\theta_{1}-\sin\theta_{2}$, which represents the reflectance of the surface. 
In our formulation, as mentioned in Sec. 2, the outgoing WDF is written as
\begin{linenomath}
\begin{equation}
R_o\left(x, u_o\right) = \int {W}_{t}\left(x, u_o; u_i\right)R\left(x, u_i\right)\textrm{d}u_i.
\end{equation}
\end{linenomath}
Assuming that a plane wave is incident on a surface ($R_{i}(x,u_{i}) = \delta(u_{i}-\sin\theta_{i}/\lambda)$) and the observer is at infinity as in the diffraction shader equations ($u_{o}=\sin\theta_{o}/\lambda$), we obtain the reflected light as
\begin{linenomath}
\begin{equation}
I(u_o) = \int {R}_o(x,u_o)\textrm{d}x
= \int {W}_t\left(x, \frac{\sin\theta_{o}}{\lambda}; \frac{\sin\theta_{i}}{\lambda}\right) \text{d}x.
\end{equation}
\end{linenomath}

In the implementation of diffraction from CDs, we assumed angle--shift invariance, where the phase function $\psi(x)$ does not depend on the incident angle. If we use the same setup assumptions as DS (i.e., a plane wave incident on a camera at sufficiently far distance), then our result is comparable to the one generated by DS, especially in the paraxial region.

The WBSDF analysis also points to new ways to improve the DS approach. DS can now be used for reflection using Eqs.~\eqref{eq:transmission} and \eqref{eq:reflection}. In the examples generated by DS, height--maps were used to model phase delays for pre--computing interference from parallel light rays, ignoring the amplitude component of the microstructure. The amplitude controls the amount of light that is absorbed or scattered. To account for the amplitude variation, we can add an $a(x)$ term in Eq.~\eqref{eq:microstructure}. Despite these extensions, DS remains applicable only in the far field, since the integration is already performed. DS is also less suitable for simulating audio diffraction (large wavelength) or simulating a PSF of a camera's optics (larger integrating cone for receiving patch). 

\subsection{Comparison with Augmented Light Fields}
Augmented light fields~\cite{Oh2009} (ALF) is the first model to use the WDF for rendering. It uses a simplistic two plane light field parametrization: one for diffraction grating plane and one for receiver plane. Hence, it is limited to demonstrations of a planar wavefront transmitted via the first plane of a diffractive occluder. They use a 'destination based' approach (i.e. differed diffraction) using an OpenGL fragment shader. The ALF is computed in a backward manner: at each point on the receiver, the shader computes the incident ray-bundle (slice of the lightfield) using an analytical formula for the diffraction pattern from the source towards the receiver. This differed diffraction is very similar to OPD, so it provides no importance sampling and the paths to trace grow exponentially with each bounce. The ALF work suggests photon mapping as a possible extension, but it is treated as a procedure that would modify the scatter stage of a renderer. By encoding a new BSDF, we make \textit{no} change to the renderer. Instead of a new rendering strategy, WBSDF encodes the microstructure into a new reflectance function independent of other elements or illumination in the scene. This allows a wide range of effects not shown with ALF such as reflection, emission, multi-bounce, importance sampling, sound rendering and a natural fit with the rendering equation. Refer to Table 1 for more details.

\subsection{Comparison with Edge--diffraction}

\begin{figure}[h!]
\centering
\includegraphics[width=\columnwidth]{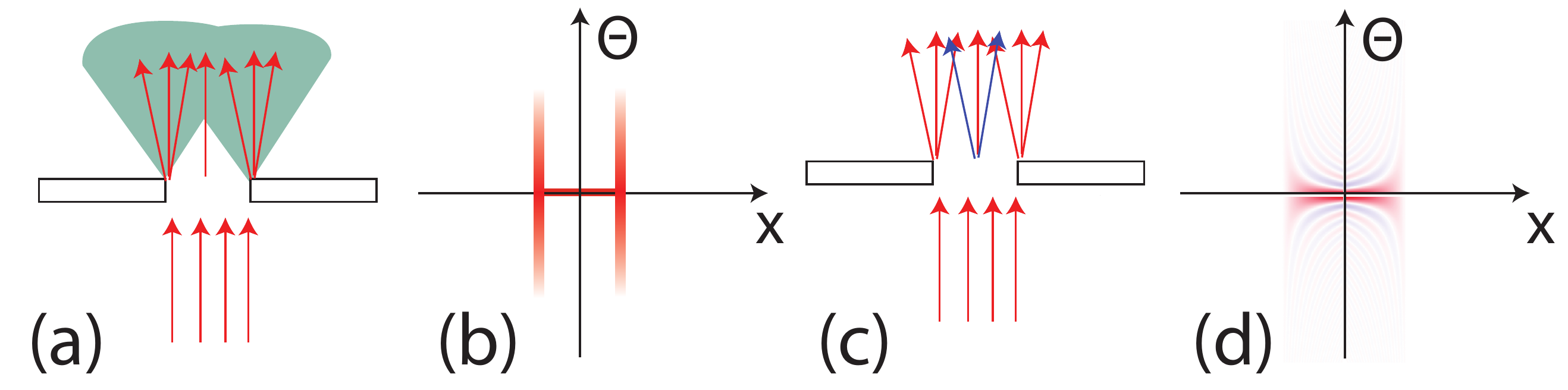}
\caption{WBSDF is more accurate than edge diffraction. (a) Edge diffraction only diffracts rays at corners and (b) ignores mid--air diffraction. (c--d) WBSDF creates midair diffraction.} 
\label{fig:edge-diff}
\end{figure}

Edge diffraction is a rudimentary form of importance sampling used in audio rendering. It can be improved by using the WBSDF as shown in Figure~\ref{fig:edge-diff}. This becomes even more significant when phase is involved. Edge diffraction techniques underestimate mid--air diffraction and allow strong ray bending at edges, whereas our technique creates diffraction through entire features, ensuring accurate diffraction results. Edge diffraction cannot achieve the unusual effect of the sound of a door closing. Our current method does not take traveling distance into account and as a result does not exhibit damping.

\section{Conclusion}
We describe a new representation of BSDF that greatly simplifies simulations of wave-phenomena in ray-based renderers. 
We bring the wave phenomena into the realm of the rendering-equation which supports global illumination, including benefits such as importance sampling. Additionally, we provide a detailed comparison of many wave phenomena renderers. We feel that we have methodically investigated the rendering of wave phenomena by proposing an easy and efficient solution, demonstrated examples in multiple domains (light/sound), investigated and solved implementation issues, and mathematically expressed out work's relationship with diffraction shaders.

%



We see many promising directions future exploration. The WBSDF can also be used for other areas of spectrum such as microwaves and x--rays or other wave--related disciplines, including the transient response for fluid wave diffraction. The simulation of camera optics can be extended to support a full camera model in existing rendering frameworks. Sound rendering extensions include attenuation, reverberation, non-planar area sources, Doppler effect and modeling of head-related transfer function (HRTF) using our formulation for easy ray-based analysis. Advanced surface scattering models, e.g., tangent--plane approximation or phase--screen approximation, will potentially extend the applications of the WBSDF and improve the accuracy to the physical models. Although current reflectance field scanning methods use a ray-based understanding, we hope WBSDF methods will inspire novel capture and inverse problems in analysis of real world objects exhibiting wave phenomena.

\section*{Acknowledgements}
Tom Cuypers, Tom Haber and Philippe Bekaert acknowledge financial support by the European Commision (FP7 IP 2020 3D Media), the European Regional Development Fund (ERDF), the Flemish Government and the Flemish Interdisciplinary Institute for Broadband Technology (IBBT). Ramesh Raskar is supported by an Alfred P. Sloan Research Fellowship. We also like to thank George Barbastathis, Markus Testorf, Roarke Horstmeyer and all our colleagues at the Camera Culture Group for their time and input.


\small{

}

\section*{Appendix}

Wigner distribution function for 1D element $t(x)$, in Matlab.
\begin{figure}[h!]
\includegraphics[width=0.8\columnwidth]{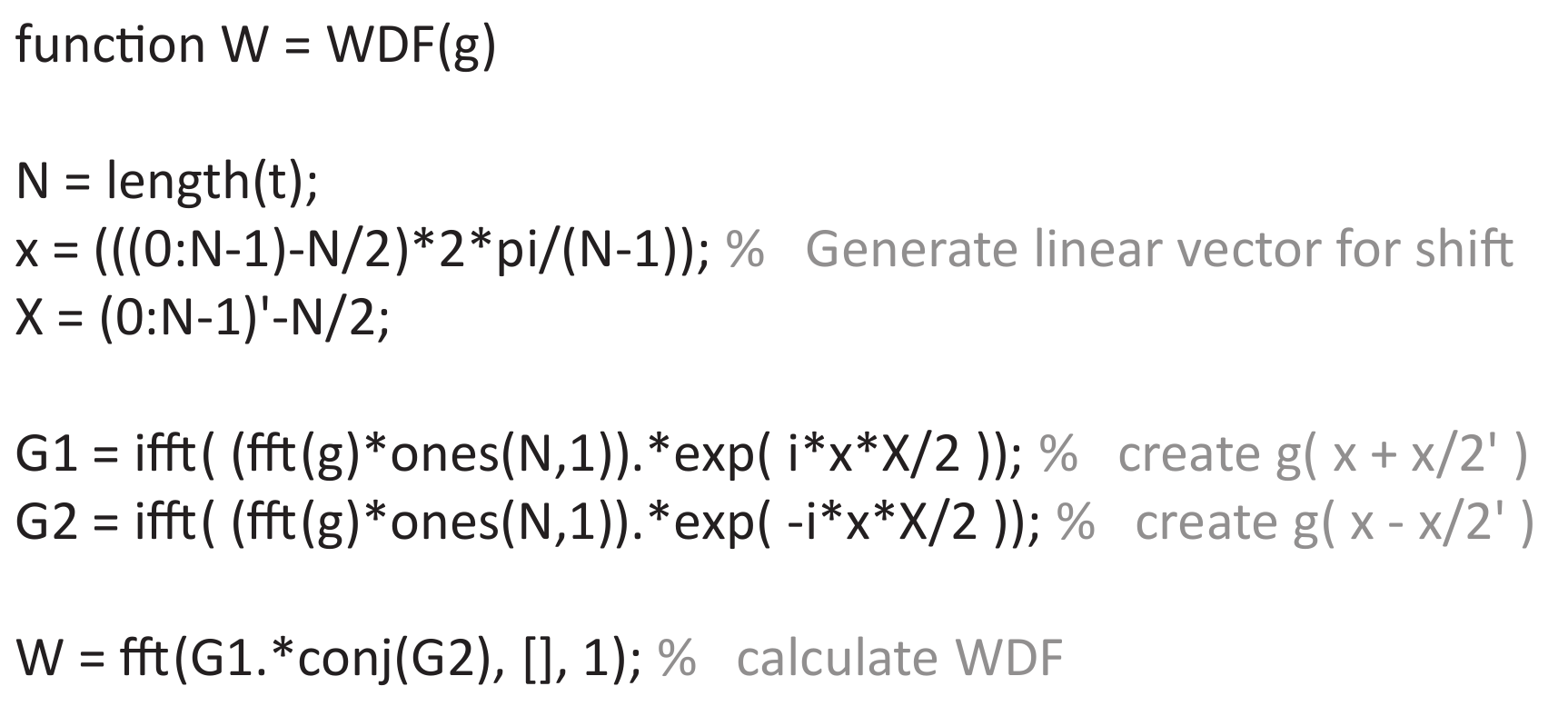}
\vspace*{-0.1cm}
\label{fig:appendixmatlab}
\end{figure}

\end{document}


\maketitle

\section{Supplemental Material}

\begin{figure}[t!]
\centering
\includegraphics[width=0.8\columnwidth]{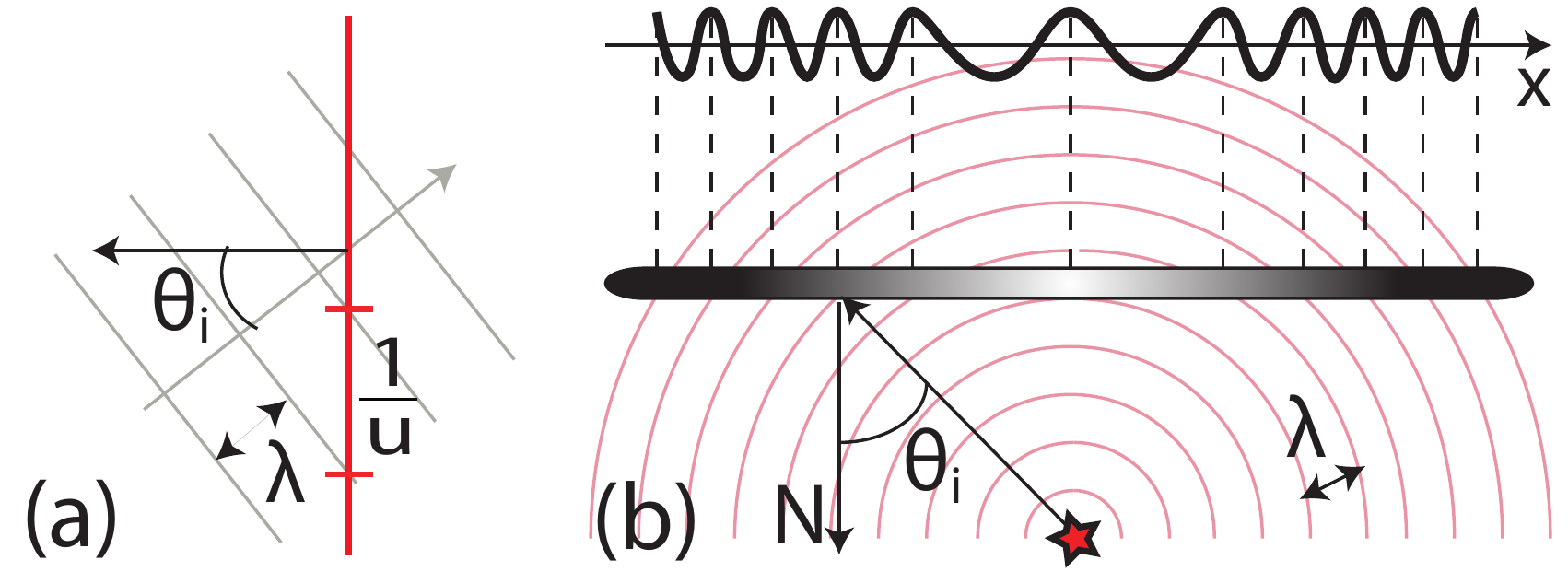}
\caption{Spatial frequency $u$ of incoming light is dependent on the incident angle $\theta_i$ and the wavelength $\lambda$ of the light. The steeper the incoming angle the higher the spatial frequency becomes.}
\label{fig:spatial_freq}
\end{figure}

\subsection{Hologram recordings}

\begin{figure}[t!]
\centering
\subfigure[Recording first hologram]{
\includegraphics[width=0.25\textwidth]{images/holo-sketch1.pdf}
}
\subfigure[Recording hologram with a slit]{
\includegraphics[width=0.3\textwidth]{images/holo-sketch2.pdf}
}
\subfigure[Recosntruction of the second hologram]{
\includegraphics[width=0.3\textwidth]{images/holo-sketch3.pdf}
}
\end{figure}

Here we present how our BRDF of rainbow holograms, which are found on many credit cards, are computed. The rainbow hologram, invented by Benton~\shortcite{Benton1969}, is reconstructed by white light and exhibits only the horizontal parallax. The recording process of the rainbow hologram consists of two steps: after the first hologram containing object information is recorded, a horizontal slit is located on top of the first hologram and reconstructed via phase conjugation, and then the second hologram captures the phase conjugated reconstruction from the first hologram~\cite{Goodman2005}. The slit allows preserving the horizontal parallax but eliminating color smearing along the vertical direction.  

Considering the desired hologram signal, we obtain the WDF of the hologram $\mathcal{W}_{h}$ as
\begin{linenomath}
\begin{multline}
\mathcal{W}_{h}(x,u,y,v) = \mathcal{W}_{obj}(x+\lambda_rz_0u, u, y+\lambda z_0 (v+v_0), v+v_0)  \\
\times\textrm{rect}\left(\frac{v+v_0}{A/(z_A\lambda_r)}\right),\qquad\qquad\qquad
\end{multline}
\end{linenomath}
where $\mathcal{W}_{obj}$ is the WDF of the object being recorded, $z_0$ is the distance between the reconstructed object and the second hologram, $z_A$ is the distance between the second hologram and the slit, $\lambda_r$ is the recording wavelength, and $v_0$ is the spatial frequency along the vertical direction of the reference wave \#2.

\subsection{Comparison with diffraction shader}

Diffraction shaders are very effective in rendering the far field effect of a single bounce. We show that diffraction shader approach is a special case of our method where the light source and a detector (camera) are at infinity. In other words, only the parallel ray bundles are considered. Based on the assumption and equations presented by Stam~\shortcite{Stam1999}, we present the relationship with WBSDF as follows.
\begin{linenomath}
\begin{multline}
I\left(ku'\right) = \psi\left(ku'\right)\psi\left(ku'\right)\\
~\quad=\int \textrm{e}^{ikwh(x)}\textrm{e}^{iku'x}\textrm{d}x \int \textrm{e}^{-ikwh(x')}\textrm{e}^{-iku'x'}\textrm{d}x'\\
=\iint \textrm{e}^{ikwh(p+\frac{q}{2})}\textrm{e}^{-ikwh(p-\frac{q}{2})}\textrm{e}^{iku'q}\textrm{d}p\textrm{d}q~~\\
=\int \mathcal{W}_{d}(x, u') \text{d}x,\qquad\qquad\qquad\qquad\qquad\quad~~
\label{eq:diff_shader}
\end{multline}
\end{linenomath}
where $k=2\pi/\lambda$, $u' = \sin\theta_1-\sin\theta_2$, $w=-\cos\theta_1-\cos\theta_2$, $p=(x+x')/2$, $q=x-x'$, and $\mathcal{W}_{d}(x,u')$ is the WDF of $\text{e}^{ikwh(x)}$ with respect to $\sin\theta_{1}-\sin\theta_{2}$, which represents the reflectance of the surface. 
In our formulation, as mentioned in Sec. 2.2, the outgoing WDF is written as
\begin{linenomath}
\begin{equation}
\mathcal{W}_2\left(x, u_2\right) = \int \mathcal{W}_t\left(x, u_2; u_1\right)\mathcal{W}_1\left(x, u_1\right)\textrm{d}u_1.
\end{equation}
\end{linenomath}
Assuming that a plane wave is incident on a surface ($\mathcal{W}_{i}(x,u_{1}) = \delta(u_{1}-\sin\theta_{1}/\lambda)$) and a detector is at infinity as in the diffraction shader equations ($u_{2}=\sin\theta_{2}/\lambda$), we obtain the reflected light as
\begin{linenomath}
\begin{equation}
I(u_2) = \int \mathcal{W}_2(x,u_2)\textrm{d}x
= \int \mathcal{W}_{t}\left(x, \frac{\sin\theta_{2}}{\lambda}; \frac{\sin\theta_{1}}{\lambda}\right) \text{d}x.
\end{equation}
\end{linenomath}
Depending on scattering models, various types of $\mathcal{W}_{t}$ are possible. In the diffraction shader, the tangent--plane approximation, where the correlation function depends on both incident and outgoing angles~\cite{Hoover2006}, has been used, and $\mathcal{W}_{d}(x, \left(\sin\theta_1-\sin\theta_2)/\lambda\right) = \mathcal{W}_{t}\left(x, \sin\theta_2/\lambda; \sin\theta_1/\lambda\right)$. 

If we assume that the angle--shift invariance for further simplification, then the phase function due to a surface can be written as $t_{h}(x) = \textrm{e}^{ik2h(x)}$ and the output WDF is
\begin{linenomath}
\begin{equation}
\mathcal{W}_2\left(x, u_2\right) = \int \mathcal{W}_t\left(x,u_1\right)\mathcal{W}_1\left(x, u_2-u_1\right)\textrm{d}u_1.
\end{equation}
\end{linenomath}
If we assume the source and camera at infinities, then 
\begin{linenomath}
\begin{equation}
I(u_2) = \int \mathcal{W}_2(x, u_2)\textrm{d}x = \int \mathcal{W}_{t_h}(x,u_2-u_1)  \textrm{d}x,
\label{eq:our_eq_diff_shader}
\end{equation}
\end{linenomath}
where $\mathcal{W}_{t_h}$ is the WDF of $\text{e}^{ik2h(x)}$ and this equation is similar to eq.~\eqref{eq:diff_shader}. The only difference is that $2h(x)$ and $wh(x)$ in the input functions, where they become identical if we use $\theta_1=\theta_2=0$. In other words, eq.~\eqref{eq:our_eq_diff_shader} would produce similar results as the diffraction shader in the paraxial region. 

As mentioned in Sec. 2. 2, a statistical model for the phase function can also be applicable to our WDF based method as in the diffraction shader. Contrast to the diffraction shader, we keep the WDF of the surface. Hence, we can use different integration kernels and approximation; e.g., reflection in the near--field can be computed and characteristics of light capturing devices such as cameras can be taken into account in reflection rendering.

For a summary, our approach is more generalized, and different models and approximations, depending on surface profile, light source and camera geometry, speed, tolerance of error, can be incorporated even in the near--field and real camera models.

\subsection{Creating BRDFs}

\subsection{Correlation Function based WBSDF}

Based on the generalized van Cittert--Zernike theorem in optics~\cite{Goodman2000}, the intensity scattered from a surface can be described by the Fourier transform of the coherence factor $\gamma(x, \Delta x)$, which is a normalized correlation function of electric--field ($\langle E(x-\Delta x/2)E^*(x+\Delta x/2)\rangle$) at the surface~\cite{Wolf1978}.

When the exact surface surfaceprofile is unknown, we can derive a WBSDF from the correlation function $\gamma(x, \Delta x)$ of the surface. Where $x$ defines \TODO{TC}{bla} and $\Delta x$ \TODO{TC}{bla2}. Since the WDF also can be defined with respect to the correlation function~\cite{Bastiaans1997}, we can derive the WBSDF from the WDF of the correlation function as 
\begin{linenomath}
\begin{equation}
\mathcal{W}_{\gamma}(x, u_{o}) = \int \gamma(x, \Delta x) \text{e}^{-i 2\pi u_{o} \cdot \Delta x}\text{d}\Delta x, \label{eq:WDF_gamma}
\end{equation}
\end{linenomath}
where $u_{o}$ is the local spatial frequency of the outgoing light. If the correlation function depends on the incident light as $\gamma(x, \Delta x; u_{i})$, then we need to compute the WDF for all the incident light as $\mathcal{W}_{\gamma}(x, u_{o}; u_{o}, u_{i})$. Then the outgoing WDF is expressed as
\begin{linenomath}
\begin{equation}
\mathcal{W}_{o}(x, u_{o}) = \int \mathcal{W}_{\gamma}(x, u_{o}; u_{i}) \mathcal{W}_{i}(x, u_{i})\text{d}u_{i}.\label{eq:WDF_uo_ui}
\end{equation}
\end{linenomath}
Equations~\eqref{eq:WDF_gamma} and~\eqref{eq:WDF_uo_ui} imply that the exact BRDF can be computed provided that the exact surface profile is known. However, it is often challenging to express the exact micro structure. Thus, it is often more convenient to take statistical average of the correlation function, related to parameters such as roughness or periodicity. This approach has been demonstrated in the diffraction shader~\cite{Stam1999} and BRDF estimation~\cite{Hoover2006}. 

the statistical properties of the structure to indicate smoothness, periodicity or roughness using an auto-correlation function and standard deviation.
Note that WDF is essentially the Fourier transform of the input signals auto-correlation function, so this is highly convenient.
Assume we can describe the surface by the autocorrelation function of the surface profile $R_h( \Delta x )$ and a standard deviation $\sigma_h$. The correlation function can be described by:
\begin{equation}
\gamma( \Delta x ) = e^{(i \frac{2 \pi}{\lambda} \textrm{sin}\theta_i \Delta x)} e^{-\left[ \frac{2 \pi}{\lambda} \sigma_h (1 + cos \theta_i)\right]^2 + \left[ \frac{2 \pi}{\lambda} \sigma_h (1 + cos \theta_i)\right]^2 R_h( \Delta x ) }
\end{equation}
Where $\theta_i$ is the incident ray angle to the surface normal.
We can derive the Wigner Distribution Function for this surface which will give us:
\begin{eqnarray}
I( u ) &\sim& \frac{cos^2 \theta_s}{\lambda^2} \int \int \gamma( \Delta x ) e^{-i k_s \bullet \Delta x} d \Delta x \\
I( u ) &\sim& \frac{ cos^2 \theta_s }{\lambda^2} e^{- \left[ \frac{2 \pi}{\lambda} \sigma_h (1 + cos \theta_i)\right]^2}\\
 & & \Im \left[ \left(\frac{2 \pi}{\lambda} (1+ cos \theta_i) \right)^2 R_h( \Delta x ) \right] |_{\Delta x \rightarrow u - \frac{sin \theta_i}{\lambda}}
\end{eqnarray}
Where $u =  \frac{\textrm{sin} \theta_s}{\lambda}$.

\subsection{Internal reflections}
\TODO{Are we keeping this section?}{}
\begin{figure}[t!]
\centering
\subfigure[Internal reflections]{
\includegraphics[width=0.45\columnwidth]{images/interrefl.pdf}
}
\subfigure[Rendering]{
\includegraphics[width=0.45\columnwidth]{images/water-oil1.pdf}
}
\caption{(a) Shows the principle of internal reflections. Light going out in the same direction are slightly out of phase due to the different travel distances. They end up at the same pixel and create interferences. We can observe this for example when a thin oilfilm is floating on top of water as rendered in (b).}
\label{fig:rec_aperture_example}
\end{figure}


\textbf{Statistically averaged WBSDF}  

Equation (6) implies that the BSDF can be computed provided the exact surface profile of the material. However, it is often challenging to express the micro structure exactly. In this situation, we can compute the WBSDF with a statistical average as
\begin{equation}
{W}_{t}(x,u) = \int \Big< t\left(x+\frac{x'}{2}\right) t^{*}\left( x-\frac{x'}{2}\right) \Big>\text{e}^{-i2\pi x' u}\text{d}x',
\end{equation}
where $\langle~\rangle$ denotes average. Depending on the surface properties and rendering environments, different types of statistical average can be used; in general, the Gaussian statistics is assumed and statistics parameters such as standard deviation or autocorrelation length can be tuned. This statistical average approach has been used in the diffraction shader~\cite{Stam1999} and BRDF estimation~\cite{Hoover2006}. Note that $\langle t\left(x+x'/2\right) t^{*}\left( x-x'/2\right) \rangle$ is sometimes referred to as the correlation function $\gamma(x, x')$. If the correlation function depends on angles of incident and/or outgoing rays, as for example in the diffraction shader, the WDF is expressed as
\begin{linenomath}
\begin{equation}
{W}_{o}(x, u_{o}) = \int {W}_{\gamma}(x, u_{o}; u_{i}) {W}_{i}(x, u_{i})\text{d}u_{i}.\label{eq:WDF_uo_ui}
\end{equation}
\end{linenomath}

\small
\bibliographystyle{acmsiggraph}
\bibliography{siggraph}